\documentclass[11pt]{article}
	
	\newcommand{\blind}{0}
	
    \makeatletter
    \renewcommand\section{\@startsection {section}{1}{\z@}%
                                       {-3.5ex \@plus -1ex \@minus -.2ex}%
                                       {2.3ex \@plus.2ex}%
                                       {\normalfont\fontfamily{phv}\fontsize{16}{19}\bfseries}}
    \renewcommand\subsection{\@startsection{subsection}{2}{\z@}%
                                         {-3.25ex\@plus -1ex \@minus -.2ex}%
                                         {1.5ex \@plus .2ex}%
                                         {\normalfont\fontfamily{phv}\fontsize{14}{17}\bfseries}}
    \renewcommand\subsubsection{\@startsection{subsubsection}{3}{\z@}%
                                        {-3.25ex\@plus -1ex \@minus -.2ex}%
                                         {1.5ex \@plus .2ex}%
                                         {\normalfont\normalsize\fontfamily{phv}\fontsize{14}{17}\selectfont}}
    \makeatother
	\usepackage{amsmath}
        \usepackage{authblk}
	\usepackage{graphicx}
	\usepackage{enumerate}
	\usepackage{natbib} %comment out if you do not have the package
	\usepackage{url} % not crucial - just used below for the URL
        \usepackage{booktabs}
        \usepackage{multirow}
        \usepackage{makecell}
        \usepackage[boxruled,vlined]{algorithm2e}
        \usepackage{tikz}
        \usepackage{bm}
        \usepackage{pgfplots}
        \usetikzlibrary{shapes}
        \usetikzlibrary{backgrounds,calc}
        \usetikzlibrary{arrows}
        \usepackage{pgfplots}
        \pgfplotsset{compat=1.8}
        \usepackage{graphicx}
        \usepgfplotslibrary{statistics}
        \usepackage{float}
        \usepackage{xcolor}

        \usepackage[caption=false]{subfig}
        \usepackage{pgfplots}
        \pgfplotsset{compat=1.18} 
        
        \usepackage{natbib}
         \bibpunct[, ]{(}{)}{,}{a}{}{,}%

        \newtheorem{theorem}{Theorem}
        \newtheorem{proposition}{Proposition}
        \newtheorem{lemma}{Lemma}
        
        \usepackage{soul}
        \usepackage[finalnew]{trackchanges}
        
        \soulregister\ref7
        \soulregister\cite7

        \addeditor{tianyu}
\begin{document}
	%%%%%%%%%%%%%%%%	
			%%%%%%%%%%%%%%%%%%%%%%%%%%%%%%%%%%%%%%%%%%%%%%%%%%%%%%%%%%%%%%%%%%%%%%%%%%%%%%
		\def\spacingset#1{\renewcommand{\baselinestretch}%
			{#1}\small\normalsize} \spacingset{1}
		%%%%%%%%%%%%%%%%%%%%%%%%%%%%%%%%%%%%%%%%%%%%%%%%%%%%%%%%%%%%%%%%%%%%%%%%%%%%%%
		
		\if0\blind
		{
			\title{Optimizing Feature Selection in Causal Inference: A Three-Stage Computational Framework for Unbiased Estimation}
			\author{
            Tianyu Yang and Md.\ Noor-E-Alam$^\ast$ 
             \\ Department of Mechanical and Industrial Engineering
             \\ Northeastern University, Boston, MA 02115, USA
                            \\$^\ast$  Corresponding author email: mnalam@neu.edu
            }
			\maketitle
				} \fi
        
% 		\ARTICLEAUTHORS{%
% \AUTHOR{Tianyu Yang}
% \AFF{Department of Mechanical and Industrial Engineering, Northeastern University, Boston, MA 02115, USA\\ \EMAIL{Email: yang.tianyu@northeastern.edu}}
% \AUTHOR{Md.\ Noor-E-Alam}
% \AFF{Department of Mechanical and Industrial Engineering, Northeastern University, Boston, MA 02115, USA \\
% \EMAIL{Email: mnalam@neu.edu}\\ \URL{Website: https://mnalam.com}}

		\if1\blind
		{

            \title{\bf {Optimizing Feature Selection in Causal Inference: A Three-Stage Computational Framework for Unbiased Estimation}}
			\author{Author information is purposely removed for double-blind review}
			
\bigskip
			\bigskip
			\bigskip
			\begin{center}
				{{Optimizing Feature Selection in Causal Inference: A Three-Stage Computational Framework for Unbiased Estimation}}
			\end{center}
			\medskip
		} \fi
		\bigskip
		
	\begin{abstract}
            Feature selection is an important but challenging task in causal inference for obtaining unbiased estimates of causal quantities. 
            Properly selected features in causal inference not only significantly reduce the time required to implement a matching algorithm but, more importantly, can also reduce the bias and variance when estimating causal quantities.
            When feature selection techniques are applied in causal inference, the crucial criterion is to select variables that, when used for matching, can achieve an unbiased and robust estimation of causal quantities. Recent research suggests that balancing only on treatment-associated variables introduces bias while balancing on spurious variables increases variance. To address this issue, we propose an enhanced three-stage framework that shows a significant improvement in selecting the desired subset of variables compared to the existing state-of-the-art feature selection framework for causal inference, resulting in lower bias and variance in estimating the causal quantity. We evaluated our proposed framework using a state-of-the-art synthetic data across various settings and observed superior performance within a feasible computation time, ensuring scalability for large-scale datasets. Finally, to demonstrate the applicability of our proposed methodology using large-scale real-world data, we evaluated an important US healthcare policy related to the opioid epidemic crisis: whether opioid use disorder has a causal relationship with suicidal behavior.\end{abstract}
			
	\noindent%
	{\it Keywords:} Causal inference, observational studies, variable selection, machine learning, adaptive elastic net.

	%\newpage
	\spacingset{1.5} % DON'T change the spacing!
%In observational studies, researchers can only observe subjects without intervention \citep{yao2021survey}.
%%%%%%Section 1%%%%%%%
\section{Introduction} \label{sec:introduction}
%%%%%%Section 1%%%%%%%
In this paper, we focus on feature selection methods in causal inference. Causal inference often deals with data obtained from observational studies, where the objective is to find the cause-effect relationship between a particular policy/intervention and the outcome.  However, identifying the cause-effect relationships could be challenging since the mechanism of how covariates affect the outcome is invisible. The most effective approach to establishing causality is through randomized controlled trials (RCTs). RCTs involve randomly assigning participants to a treatment or control group and subsequently comparing outcomes. With a sufficiently large dataset, random assignment ensures covariate balance between the two groups. The difference between the expectation of the outcome predictors in the two groups provides insights into the causality of the treatment indicator and outcome predictor. Nevertheless, Randomized Controlled Trials (RCTs) are often time-consuming and expensive \citep{yao2021survey}. Consequently, matching methods leveraging observational data are frequently employed. These methods aim to achieve covariate balance by equalizing the distribution of covariates in the treated and control groups \citep{stuart2010matching}. Regardless of the chosen matching algorithm, feature selection plays an important role not only in significantly reducing the time required to execute subsequent matching algorithms but also in aiding the attainment of unbiased and robust estimates of causal quantities.

% %%%%%%Section 1.1%%%%%%%
% \subsection{Role of feature selection in causal inference}
% %%%%%%Section 1.1%%%%%%%

As a motivating example, assume a pharma company is investigating the potential positive impact of their new medicine on treating heart disease in patients. For this study, they selected a sample of 1,000 individuals from a population of 1,000,000 and documented the observed effects. The company collected data on more than 30 covariates for each of the 1,000,000 individuals, encompassing factors such as age, gender, family history of heart disease, etc. 

When considering the feature selection process for the above problem, one might initially use traditional feature selection tools (e.g., lasso- or ridge-based models). Subsequently, based on the selected features, one may calculate the difference between the predicted outcomes of the treated and control groups and claim this difference as the estimated causal quantity. However, recent studies suggest that traditional feature selection tools, when used directly, often lack adequate performance in estimating the unbiased estimation of causal quantity \citep{stuart2010matching, rotnitzky2010note}. Consequently, the estimation of the causal effect lacks robustness and reliability. \citep{ho2017om}.

To explain this, first, we denote the observation for a person $i$ as $Y_i=Y_i(T=1)\bm{1}(T=1)+Y_i(T=0)\bm{1}(T=0)$. Let's assume person $i$ is among the 1,000 individuals who received treatment, thus $Y_i(T=1)$ is observed. However, $Y_i(T=0)$ remains unobserved. This unobserved outcome for each person is referred to as the counterfactual outcome. Because of the existence of counterfactual outcomes, if the pharmaceutical company selects the treatment group based on another factor that can also influence the outcome, such as \textit{age}, then it becomes challenging to confidently determine whether the observed effect is attributable to the new medicine or the variable \textit{age}.

Some recent studies also emphasize the necessity of feature selection of causal inference. For example, propensity score matching suffers from variance inflation caused by incorporating unnecessary covariates \citep{de2011covariate}. Additionally, incorporating all confounders for unbiased treatment estimates, while simultaneously eliminating extraneous variables, leads to lower bias and variance in the estimation of causal quantities \citep{shortreed2017outcome, greenland2008invited, rotnitzky2010note}.

Feature selection in causal inference is crucial for drawing robust conclusions about unbiased causal quantities. A key standard for achieving robust conclusions in causal inference is selecting the maximum number of matched pairs \citep{islam2019robust}. Consider a scenario in which one-to-one matching is applied in the pharmaceutical company example, selecting up to 600 matched pairs based on three variables: age, gender, and family history of heart disease. However, if only two variables—age and gender—truly influence the causal quantity, matching solely on these two variables could yield up to 1,000 matched pairs. According to the theory proposed by \cite{islam2019robust}, matching on three variables selects fewer matched pairs than 1,000 and violate the \textit{absolute robust} rule, meaning the \textit{p}-values of the maximum and minimum test statistics for the three variables are unequal. Consequently, matching on a higher-dimensional subset of variables than necessary increases bias and variance in causal inference.

% %%%%%% Section 1.2 %%%%%%%
% \subsection{Selection bias and variance in causal inference}
% %%%%%% Section 1.2 %%%%%%%
We now discuss which features should be included to estimate the causal quantity, and then provide an insightful discussion on the feature selection process in causal inference.

Recent studies show that the best subset of the variables to be selected should include confounders (covariates that are both associated with the treatment indicator and the outcome predictor) and pure outcome predictors (variables that are only associated with the outcome predictor)\citep{islam2021feature, shortreed2017outcome}. In other words, the efficient variable selection method should exclude pure treatment predictors (variables only associated with the treatment predictors) and noise (variables that are unassociated with either treatment predictors or outcome predictors). 

For simplicity in use, we denote the pure treatment predictors as $\bm{X_T}$, the confounders as $\bm{X_C}$, the pure outcome predictors as $\bm{X_P}$, the noise covariates as $\bm{X_S}$.

Let us assume $Y$ is the outcome predictor, $T$ is the treatment indicator, and $T=\{0, 1\}$, $\epsilon$ is the noise. \change[tianyu]{The causal quantity we use is called the average treatment effect on the treated (ATT)}{We utilize the average treatment effect on the treated (ATT) as the causal quantity}, which is defined as $E[Y(T=1)|T=1] - E[Y(T=0) | T=1]$. The first term refers to the expectation of the outcome when the treated units receive treatment, and it can be calculated by averaging the outcome across all the treated units. The second term, confoundedness, refers to the expected outcome when treated units do not receive treatment. Since this situation is unobserved, the second term is unknown to us.

\begin{lemma}[Selection Bias and Variance of Estimation of the Causal Quantity] \label{lemma: ubest}
    Assuming the outcome model is the linear combination of the treatment indicator and the subset of selected variables, i.e., $\bm{Y}=\bm{T}\alpha+\bm{X\beta}+\bm{\epsilon}$. To achieve unbiased estimation of causal quantity, the model should select $\bm{X_C}$ and $\bm{X_P}$, and exclude $\bm{X_T}$. Additionally, including $\bm{X_S}$ will introduce variance to the estimation of the causal quantity.
\end{lemma}

Serving as one of the primary motivations for this paper, we provide detailed proofs \citep{wooldridge2016should, pearl2012class} that are widely accepted for the proof of Lemma \ref{lemma: ubest} \citep{vanderweele2019principles, yao2021survey, shortreed2017outcome, islam2021feature}. The proofs are available in the \textbf{Online Supplement S1 Appendix A}.\\
%%% may need to add proofs from \citep{pearl2012class} as well %%% 

The lemma \ref{lemma: ubest} indicates the following proposition:

\begin{proposition} [Ideal Subset of Variables to Include] \label{prop:ideal_var_sel}
    The efficient variable selection method should include $\bm{X_C}$ and $\bm{X_P}$ while exclude $\bm{X_T}$ and $\bm{X_S}$ in the feature selection process.
\end{proposition} 

Another motivation for this paper is the need for the oracle property in feature selection when working with large-scale datasets. The oracle property emphasizes the consistency of the set of variables chosen during the feature selection process. For example, in studying the effect of the treatment \textit{high blood pressure} on the outcome \textit{diabetes}, the selected set of features may not consistently include \textit{BMI} across different seeds or datasets sampled from the raw data. This inconsistency can lead researchers to question whether \textit{BMI} should be included in the matching process. For many matching algorithms, the inclusion or exclusion of such features can significantly impact the bias, variance, and computational time involved in estimating causal quantities.

%%%%%%Section 2%%%%%%%
\section{Literature Review}
%%%%%%Section 2%%%%%%%

In this section, we focus on the feature selection of both traditional and state-of-the-art causal inference literature. Ideally, exact matching would be the best way to reach an unbiased estimation of causal quantity since it reaches the exact covariate balance. However, strict exact matching may result in no matched pairs when dealing with high-dimensional data. Therefore, various distance metrics are developed for matching process. For instance, Mahalanobis Distance Matching (MDM) defines Mahalanobis distance \cite{rubin1979using}, while Coarsened Exact Matching (CEM) \cite{iacus2012causal} introduces binning methods. However, calculating Mahalanobis distance is time-consuming \citep{stuart2010matching}, while the strata of CEM required for matching would increase exponentially \citep{roberts2015matching}. Propensity score matching (PSM), introduced by \cite{rosenbaum1983central}, is more popular than previous distance measures due to its efficiency and ability to handle high-dimensional data. PSM uses a propensity score defined as the probability of the units being treated as the distance measure. It maps all the covariates into a one-dimensional scalar, thus making all the points from the treatment and control groups comparable. However, some researchers argue that PSM creates a fully randomized design instead of a fully blocked randomized experimental design \citep{king2019propensity}, and may provide non-robust matched pairs.

The unscalable and non-robustness properties of traditional matching algorithms drive researchers to consider feature selection in causal inference. By reducing the dimension of the observational data, the subsequent matching algorithms can reach a lower bias and variance. \cite{wang2012bayesian} proposed the Bayesian Effect Estimation (BAC), which leverages a two-stage framework to select the features. It incorporates both prior knowledge about the confounding factors and the observed data and uses the estimation of the probability distribution of the exposure effect on the outcome model to adjust for the confounding factors. \cite{talbot2015bayesian} proposed the Bayesian Causal Effect Estimation Algorithm (BCEE), which first includes potential confounders and modify the effect of the treatment, then estimates the model parameters by using Markov Chain Monte Carlo (MCMC) methods to generate posterior distributions of the treatment effect. Although powerful for reaching unbiased estimation, BCEE is time-consuming and is not scalable to high-dimensional data. \cite{ertefaie2018variable} proposed the penalized modified objective function estimators (PMOE) method, which defines a penalized objective function that combines both least squares and maximum likelihood estimates of the parameters in the exposure and outcome model and performed slightly better than BAC.

Inheriting the two-stage structure model, \cite{shortreed2017outcome} proposed the Outcome-Adaptive Lasso (OAL) method, which leverages the principles of the adaptive lasso (ADL) initially proposed by \cite{zou2005regularization} and aligns the exposure model with the prior knowledge indicated by the coefficients of the outcome model. Following the augmented form of elastic net estimator \citep{zou2006adaptive}, which is more stable than the adaptive lasso on highly correlated datasets, \cite{balde2022reader} proposed a generalized OAL (GOAL) model for causal inference and presents a smaller bias on high-dimensional data compared to the OAL. Then, \cite{islam2021feature} proposed the outcome adaptive elastic net (OAENet) method, presenting the adaptive elastic net estimator and demonstrated through testing on both synthetic and real-world datasets, particularly showcasing its effectiveness in handling highly correlated structure datasets. However, OAENet struggles to maintain the oracle property.

In addition to the two-stage framework, several other recently developed methods based on causal graph theory have also demonstrated promising results in feature selection for causal inference. For instance, \cite{gruber2010application} proposed the collaborative targeted maximum likelihood estimation (CTMLE) method, which employs a targeted semi-parametric double robust plug-in estimator to estimate the average treatment effect (ATE). \cite{kursa2010boruta} introduced the Boruta method, which utilizes random forest for classification. In Boruta, the treatment indicator (T) and the outcome variable (Y) are separately synthesized in a random forest classifier, and the Z score is considered to minimize mean accuracy loss. \cite{de2011covariate} proposed the DWR (De Luna, Waernbaum, and Richardson) causal-graph framework, which performs non-parametric graph-based covariate selection.

Besides the feature selection methods discussed above, some researchers have also attempted to utilize machine learning based feature selection methods. For example, \cite{wang2021flame} proposed the Fast Large-scale Almost Matching Exactly Approach (FLAME) method, which leverages two factors (prediction error (PE) and balance factor (BF)) and uses a penalty hyperparameter to trade off the two standards. Then FLAME performs a backward feature selection process, aiming to choose exactly the set of covariates that can predict the outcomes. Although powerful and time-efficient, FLAME is not recommended to be used when there are continuous covariates unless the assumption that binning the covariates wouldn't affect causal estimation is met \citep{wang2021flame}.

In this paper, we propose a novel feature selection framework for causal inference that addresses the limitations of existing models. Designed to minimize selection bias and variance, the framework integrates an SVM estimator for the exposure model and a penalty smoothing function to enhance performance and achieve the oracle property. This robust three-stage approach demonstrates superior capability in identifying confounders and pure outcome predictors.

Our key contributions can be summarized as follows:

1. \textbf{Mitigate Selection Bias}: Selection bias in Causal Inference is a key challenge \citep{vanderweele2019principles, wooldridge2016should, lu2020feature}. Our proposed feature selection method is designed to overcome the selection bias. Experimental results using state-of-the-art algorithms leveraging widely used synthetic datasets \citep{shortreed2017outcome, islam2021feature, wang2012bayesian} where the ground truth (i.e., the true treatment effect) is known show that applying our proposed feature selection algorithm in the design stage helps to mitigate the selection bias of treatment estimation.
\\ \hspace*{1em} 2. \textbf{Practical Application}: Interpretability is one of the key challenges for any causal inference method. Practitioners, such as medical experts, policymakers and scientists, do not accept findings unless they are interpretable. By allowing feature selection in the design stage using the proposed technique, we enhance the robustness of causal conclusions with fewer features. The smaller set of selected features improves the interpretability of these conclusions, making them more accessible to practitioners, including policymakers and medical experts. Moreover, our proposed feature selection method selects very few features from the NSDUH dataset to estimate ATT. The insight gained from the feature selection could guide researchers to focus on fewer features while collecting data (only the selected features) for future studies of similar problems, thereby reducing time and cost in scientific research and real-world applications. 
\\ \hspace*{1em}3. \textbf{Stability and Consistency}: Real-world observational data, such as healthcare data, often suffers from class imbalance issues (where the treatment group is much smaller than the control group). Our proposed algorithm is stable and robust, especially when large number of bootstrap iterations are required to address the class imbalance issue. Experiments with a real-world healthcare dataset (NSDUH dataset) demonstrate that our proposed feature selection technique outperforms other state-of-the-art methods, such as OAENet, in maintaining consistency (oracle property) when selecting features. Our analysis illustrates the robustness of our framework, showing that the selected variables remain largely consistent across different instances drawn from the same dataset.
\\ \hspace*{1em} 4. \textbf{CPU Time}: Our feature selection algorithm improves the robustness of conclusions by reducing bias in the estimation of the average treatment effect with selected features, while also reducing computation time. This shows the scalability and promise of our proposed method for unbiased estimation of ATT leveraging real-world high-dimensional observational data.

%%%%%%%%%%%%%%Section 3%%%%%%%%%%%%%%
\section{Two-stage Feature Selection Framework in Causal inference} \label{sec:ci_relation}
%%%%%%%%%%%%%%Section 3%%%%%%%%%%%%%%
This section introduces the state-of-the-art two-stage feature selection frameworks that mitigate selection bias and variance in causal inference. 

Before introducing the two-stage framework, we first highlight the advantages of adopting a multi-stage feature selection framework over incorporating feature selection directly into the model construction process. The primary advantage lies in achieving oracle property. Multi-stage frameworks have been shown to be effective in attaining oracle property easily \citep{zou2006adaptive, zou2009adaptive, shortreed2017outcome, islam2021feature}. In contrast, within our reach, no current literature provides evidence that oracle property can be reliably achieved through a one-stage process.

The two-stage framework is composed of at least two parametric models: One is the \textit{outcome model}, and the other is the \textit{exposure model} \citep{ertefaie2018variable}. The key idea of the two-stage framework is to input the prior knowledge of covariates generated from one model to the regularization term of the other model in the feature selection process. The framework generally has the following two stages:

1. Fit the parametric exposure/outcome model and retrieve the coefficient for each covariate.

2. Map the coefficients of covariates to a parameter that gives the prior knowledge in the other model and utilize this prior knowledge to regulate the feature selection process.

In the second stage, the method of mapping the coefficient of each covariate to prior knowledge varies in different models. For instance, Bayesian Adjustment for Confounding (BAC) \citep{wang2012bayesian} and Bayesian Causal Effect Estimation (BCEE) \citep{talbot2015bayesian} initially fit the exposure model and then use the model's coefficients to specify two posterior distributions in the outcome model. On the other hand, Outcome Adaptive Lasso (OAL) \citep{shortreed2017outcome} and Outcome Adaptive Elastic Net (OAENet) \citep{islam2021feature} first fit the outcome model through regression and estimate the coefficients using the maximum likelihood method. Subsequently, they fit the exposure model based on the logistic regression model. The coefficient of each covariate in the outcome model in OAL and OAENet is interpreted as a weight parameter that helps regulate the feature selection process in the exposure model. Both OAL and OAENet are derived from lasso-based feature selection methods. The underlying concepts are thoroughly explained in \textbf{Online Supplement S1 Appendix B}.

We categorize the state-of-the-art feature selection models for causal inference into the following two types of frameworks. The two types of frameworks are distinguished based on which model we fit first.

\textbf{Framework A}: Fit the exposure model first, which utilizes the treatment label $\bm{T}$ as the response variable. Then fit the outcome model, which utilizes the outcome label $\bm{Y}$ as the response variable. For example, BAC and BCEE.

\textbf{Framework B}: Fit the outcome model first, which utilizes the outcome label $\bm{Y}$ as the response variable. Then fit the exposure model, which utilizes the treatment label $\bm{T}$ as the response variable. For example, OAL and OAENet.

In the rest of the section, we first present the feature selection model of OAL and OAENet
as our proposed framework is similar to these two models. Then, we discuss the properties of the two types of feature selection framework. In the last of this section, we will introduce our SVM-based feature selection model and its advantages.

%%%%%%%%%%%%%%%section 3.1%%%%%%%%%%%%%%%%%%
\subsection{Outcome Adaptive Lasso (OAL)} \label{sec:oal}
%%%%%%%%%%%%%%%section 3.1%%%%%%%%%%%%%%%%%%
In 2017, Shortreed et al. \cite{shortreed2017outcome} introduced the Outcome-Adaptive Lasso (OAL) method for the feature selection process in causal inference. The procedure is outlined as follows:

\texttt{outcome model: 
\begin{equation}
    \bm{\widetilde{\theta}(OLS)} = argmin_{\bm{\theta}}  \bm{||Y-X \theta||_2^2} 
\end{equation}}

\texttt{exposure model: 
\begin{equation}
    \bm{\hat{\beta}(OAL)} = argmin_{\bm{\beta}} \bigg\{\sum_{i=1}^n( -t_i(\bm{X_i^T\beta}) + log(1+e^{\bm{X_i^T\beta}})) + \lambda_n \sum_{j=1}^{p}{\hat{w}_j|\beta_j|}  \bigg\}
\end{equation}}

Here $t_i$ is the indicator for the unit $i$ being treated or not, $\hat{w}_j=|\widetilde{\theta}(OLS)_j|^{-\gamma}$, and $\bm{\hat{\beta}(OAL)}$ is the vector which indicates the coefficient assigned to each covariate. \change[tianyu]{If $\beta_j$ is a non-zero coefficient, the covariate will be selected.}{Non-zero coefficient of $\beta_j$ indicates the variable is selected. }

Intuitively, \change[tianyu]{if we have large}{large} $\widetilde{\theta}(OLS)_j$ in the outcome model lead to small \add[tianyu]{value of} $\hat{w_j}$ \remove[tianyu]{input to the exposure model}, which means, this covariate is more likely to be chosen in the exposure model. 

Outcome adaptive lasso \change[tianyu]{performs excellently in selecting a good-quality subset of variables in a computationally feasible time}{is very efficient in choosing a high-quality subset of variables suggested by Proposition 1}. \change[tianyu]{\protect However, since the method used adaptive lasso as the feature selection process, it inherits the disadvantages from the original adaptive lasso, for example, the instability in the high-dimensional data \citep{zou2009adaptive}. Analysis and experiments from \cite{islam2021two} also agree with the conclusion that outcome adaptive lasso will overly penalize the pure outcome predictors in the synthetic dataset.}{\protect Nevertheless, as this method employs adaptive lasso for feature selection, it inherits certain drawbacks from the original adaptive lasso model, including instability in high-dimensional data \citep{zou2009adaptive}. Further analysis and experiments also support the observation that outcome adaptive lasso tends to excessively penalize pure outcome predictors in synthetic datasets \citep{islam2021feature}.} 

To meet the oracle property, \cite{shortreed2017outcome} introduces a method called weighted absolute mean difference (wAMD), which defines an objective function regarding the choice of $\lambda$ that minimizes the Inverse Probability Treatment Weighting (IPTW) estimator. The related proof for the oracle property is presented in \textbf{Online Supplement S1 Appendix C.1.}

%%%%%%%%%%%%%%%Section 3.2%%%%%%%%%%%%%%%%%%
\subsection{Outcome Adaptive Elastic Net (OAENet)} \label{sec:OAENet}
%%%%%%%%%%%%%%%Section 3.2%%%%%%%%%%%%%%%%%%
\cite{islam2021feature} proposed the Outcome Adaptive Elastic Net method (OAENet)\change[tianyu]{, the method goes as below:}{. The model is formulated below:}

\texttt{outcome model: 
\begin{equation}
    \bm{\widetilde{\theta}(OLS)} = {argmin}_{\bm{\theta}}  \bm{||Y-X \theta||_2^2} 
\end{equation}}

\texttt{exposure model: 
\begin{equation}
    \bm{\hat{\beta}(OAENet)} = (1+\frac{\lambda_2}{n}){argmin}_{\bm{\beta}} \bigg\{ \sum_{i=1}^n (-t_i(\bm{X_i^T\beta}) + log(1+e^{\bm{X_i^T\beta}})) + \lambda_2 ||\bm{\beta}||_2^2+ \lambda_1 \sum_{j=1}^{p}{\hat{w}_j|\beta_j|}  \bigg\}
\end{equation}}

Compared to the OAL method, OAENet has a relatively smaller absolute bias and standard error on the ATT estimation when data is highly correlated. The outcome adaptive elastic net method also enjoys the oracle property. We also provide simple proof for this property in \textbf{Online Supplement S1 Appendix C.2.}

%%%%%%%%%%%%%%%Section 3.3%%%%%%%%%%%%%%%%%%
\subsection{Two types of two-stage framework} \label{sec:discussion}
%%%%%%%%%%%%%%%Section 3.3%%%%%%%%%%%%%%%%%%
Based on the discussion above, OAL and OAENet belong to the design of Framework B. Now we consider the possible model that applies similar ideas of OAL and OAENet to Framework A. Recalling Lemma \ref{lemma: ubest} and Proposition \ref{prop:ideal_var_sel}, only confounders and pure outcome predictors (i.e., $\bm{X_C} \And \bm{X_P}$) should be selected. Therefore, we have the following conclusions:

\textbf{Framework A}: If we fit the exposure model first, the covariates with high coefficient are composed of $\bm{X_C}$ and $\bm{X_T}$. Lemma \ref{lemma: ubest} indicates that we should exclude $\bm{X_T}$ while choosing $\bm{X_P}$ in the outcome model. Therefore, the regularization term exists only in the outcome model.  

\textbf{Framework B}: If we fit the outcome model first, the covariates with high coefficient are composed of $\bm{X_P}$ and $\bm{X_C}$. In some cases, part/all covariates of $\bm{X_T}$ might be assigned with high coefficients since it indirectly impacts $\bm{Y}$ through $\bm{T}$. Lemma 1 indicates that the exposure model should identify $\bm{X_C}$ and exclude $\bm{X_T}$. Therefore, the regularization term exists only in the exposure model. 

OAL and OAENet utilize a penalty factor $\bm{\hat{w}}=\{\hat{w}_1, \hat{w}_2, ..., \hat{w}_j\}$, where $\hat{w}_j=1/|\hat{\beta}|$ because they would like to keep $\bm{X_P}$ and $\bm{X_C}$ in the second model.

Similar to OAL and OAENet, we now apply a penalty factor $\bm{\hat{w}}=\{\hat{w}_1, \hat{w}_2, ..., \hat{w}_j\}$ to Framework A and use $|\hat{\beta}|$ to indicate the prior knowledge of the exposure model. In Framework A, we set $\hat{w}_j \propto {|\hat{\beta}|}$, as we wish to exclude $\bm{X_T}$ by applying a large penalty to $\bm{X_T}$ in the outcome model.

Note that, the regularization term exists only in the latter model in both frameworks. The regularization term will assign zero coefficients to some covariates, leading to extremely low penalties in Framework A. As zero coefficients might be assigned to noise variable $\bm{X_S}$ in framework A, consequently, the framework will select $\bm{X_S}$, which is not desired. On the other hand, zero coefficients indicate extremely large penalties in Framework B and zero coefficients might be assigned to those $\bm{X_C}$ which have relatively small associations with the outcome $\bm{Y}$. Therefore, applying the regularization term in Framework B will risk excluding some $\bm{X_C}$, which is not desired.

Our proposed method follows Framework A. In the rest part of this paper, we will delve deeper into the comparison of the two frameworks and provide justifiable explanations for our preference of Framework A over Framework B in our proposed method.

%%%%%%%%%%%%%%%Section 3.4%%%%%%%%%%%%%%%%%%
\subsection{SVM: Alternative choice for exposure model} \label{sec:svm}
%%%%%%%%%%%%%%%Section 3.4%%%%%%%%%%%%%%%%%%
In this section, we discuss the advantages of utilizing the support vector machine (SVM) to generate prior knowledge in framework A.

In the exposure model, the treatment indicator $T_i=\{0,1\}$. Assume the label of the outcome in SVM is $L$. Then, it is natural to predict $T_i=1$ if the label $L_i=1$ and predict $T_i=0$ if the label $L_i=-1$. Then the objective function is:

\begin{equation} \label{equ:svm}
    \begin{aligned}
    & \min_{\bm{\beta},b}\frac{1}{2}||\bm{\beta}||^2 \\
    & s.t ~~ L_i(\bm{\beta^T\phi(X)+b}) \leq 1 \\
    & i=1,...,n
    \end{aligned}
\end{equation}

Here, $\bm{\phi(X)}$ refers to the characteristic space and $\bm{\beta}$ refers to the estimation of the coefficient for each variable in the SVM model. Rewritten the equation we have:

\begin{equation} \label{equ:pen_svm}
\min_{\bm{\beta},b}\frac{1}{2}||\bm{\beta}||^2+C\sum_{i=1}^n max(0, L_i(\bm{\beta^T X_i} + b_i)-1)
\end{equation}

Here, $C$ is a positive constant referring to the penalty factor of misclassification. By simply observing Eq. (\ref{equ:pen_svm}), the SVM model can prevent the absolute value of coefficients for some dominant covariates from becoming extremely high (otherwise, the first term in Eq. (\ref{equ:pen_svm}) would be extremely large). In the exposure model, when using logistic regression, there is a risk of overfitting \citep{balde2022reader} when dealing with high-dimensional data, as it may assign extremely large absolute coefficients to dominant covariates. In contrast, SVM tends to assign smaller absolute weights to each covariate, which is supported by experiments on our synthetic datasets.

Now let us consider $\hat{\beta}_j$, which refers to the weight assigned to covariate $x_j$ in the exposure model. If $|\hat{\beta}_j|$ is large, and considering that in Framework A, $\hat{w}_j \propto |\hat{\beta}_j|$, the penalty assigned to this covariate in the second model will also be large, making it less likely to be chosen. Since SVM tends to assign small weights to the covariates, SVM does not separate covariates far away, this intuitively provides a relatively moderate prior knowledge than logistic regression and increases the chance of selecting $\bm{X_C}$.

Moreover, if we choose $\gamma=1$ and let $\hat{w}_j=|\hat{\beta}_j|$ in the Framework A, the expectation of $\sum_{i=1}^p\hat{w}_j^2$ is exactly $\sum_{i=1}^p\hat{\beta}_j^2$. Recall that in Eq.(\ref{equ:svm}), the objective function is to minimize $\frac{1}{2}||\bm{\beta}||_2^2=\frac{1}{2}\sum_{i=1}^p\hat{\beta}_j^2$. From the mathematical form of the elastic net estimator given by the paper \citep{zou2009adaptive} (the detailed theory and proof is presented in \textbf{Online Supplement S1 Appendix F}), SVM elastic net estimator will minimize the upper bound of $E||\bm{\hat{\beta}(\lambda_2,\lambda_1)-\beta^*}||_2^2$. The previous discussion indicates that the SVM-based elastic net estimator indicates stronger oracle property in choosing variables than logistic regression-based models.

%%%%%%%%%%%%%%%Section 4%%%%%%%%%%%%%%%%%%
\section{Two Preliminary Two-Stage Variable Selection Framework for Causal Inference} \label{sec:prop_two_stage}
%%%%%%%%%%%%%%%Section 4%%%%%%%%%%%%%%%%%%
In this section, we present the following proposed preliminary two-stage models developed based on Framework A, i.e., the models fit the exposure model first and then fit the outcome model. The preliminary framework is outlined below.

\texttt{exposure model: 
\begin{equation}
    \hat{\bm{\beta}} = \hat{\bm{\beta}}(SVM) ~ or ~ \hat{\bm{\beta}}(LR)
\end{equation}
} 

\texttt{outcome model: 
\begin{equation} \label{model:two-stage-outcome-model}
    \bm{\hat{\theta}} =(1+\frac{\lambda_2}{n})\bigg\{argmin_{\bm{\theta}}||\bm{Y}-\bm{X\theta}||_2^2 + \lambda_2 ||\bm{\theta}||_2^2+ \lambda_1 \sum_{j=1}^{p}{\hat{w}_j|\theta_j|}  \bigg\}
\end{equation}}

Here, $\hat{w}_j=\big\{ \frac{1}{1+exp(-|\hat{\beta}_j|)} \big\}^{\gamma}$, which is known as the sigmoid function with power $\gamma$. In the exposure model, the prediction label is the treatment indicator $T$. If $\hat{\bm{\beta}}(SVM)$ is adopted, then $\bm{\hat{\beta}}$ is the optimal solution of Eq.(\ref{equ:pen_svm}); If $\hat{\bm{\beta}}(LR)$ is adopted, then $\bm{\hat{\beta}}$ is the logistic regression estimator of coefficients.

For simplicity, we denote the preliminary models as \texttt{ESVMS/ELRS}. Here, \texttt{E} refers to the exposure model being fit in the first stage, SVM/LR is the estimator used in the first stage, and the last \texttt{S} refers to the sigmoid function used to smooth the penalty.

The remainder of this section will discuss the choice of parameters and compare the proposed models to previous state-of-the-art two-stage feature selection models for causal inference.

%%%%%%%%%%%%%%%Section 4.1%%%%%%%%%%%%%%%%%%
\subsection{Selecting $\lambda$} \label{sec:lambda_Sel}
%%%%%%%%%%%%%%%Section 4.1%%%%%%%%%%%%%%%%%%

Compared to Framework B, $\lambda$ in Framework A is straightforward. In the second stage, the optimal $\lambda$ should yield the most regularized model such that the cross-validated error is within one standard error of the minimum. Since we choose the adaptive elastic net as the regularization method in the outcome model, the same method presented in \cite{zou2006adaptive} could be adopted, which is to first grid and fix $\lambda_2$, then choose $\lambda_1$ so that the best pair $(\lambda_2, \lambda_1)$ minimizes the cross-validation error for the outcome model.

%%%%%%%%%%%%%%%Section 4.2%%%%%%%%%%%%%%%%%%
\subsection{Selecting $w$} \label{sec:sel_w}
%%%%%%%%%%%%%%%Section 4.2%%%%%%%%%%%%%%%%%%
Recall conclusions in section \ref{sec:ci_relation}, in Framework A, $\hat{w}_j \propto |\hat{\beta}_j|$. However, directly selecting $\hat{w}_j=|\hat{\beta}_j|$ is inappropriate. For example, when considering the noise covariates (i.e., $\bm{X_S}$) in the exposure model, if $\hat{w}_j=|\hat{\beta}_j|$, then the weights $\hat{w}_j$ of these covariates will be close to 0. When mapping $\hat{w}_j$ to the penalty value that carries prior knowledge to the outcome model, small penalties will be assigned to these variables. Therefore, it does not improve the ability to eliminate $\bm{X_S}$. To address this issue, we propose using the sigmoid function to smooth the penalty. We refer to such functions as \textit{penalty smoothing functions}.

 \begin{equation} \label{equ:sig_beta}
    \hat{w}_j=\frac{1}{1+exp(-|\hat{\beta}_j|)}
\end{equation}

Since $|\hat{\beta}_j|$ ranges from [0, $\infty$), Eq.(\ref{equ:sig_beta}) ranges from [0.5, 1). Employing the sigmoid function yields the following advantages:

1. \change[tianyu]{Sigmoid function will lift the small weights to a larger value while restricting the large weights to 1. Lifting the small weights is equivalent to lifting the penalization of $\bm{X_S}$. Restricting the large weights to 1 is equivalent to restricting the penalization of $\bm{X_C}$ and $\bm{X_T}$ to the same amount of penalty. However, the outcome model will penalize $\bm{X_T}$ and keep $\bm{X_C}$. Thus it will potentially improve the results in both eliminating $\bm{X_S} \cup \bm{X_T}$ and keeping $\bm{X_C}$.}
{The sigmoid function elevates small weights to larger values while constraining large weights to 1. Elevating the small weights is equivalent to increasing the penalization of $\bm{X_S}$. Constraining the large weights to 1 is equivalent to restricting the penalization of $\bm{X_C}$ and $\bm{X_T}$ to the same level. However, the outcome model will penalize $\bm{X_T}$ and retain $\bm{X_C}$. Therefore, this approach has the potential to improve the ability of this framework to both eliminate $\bm{X_S}$ and $\bm{X_T}$, and retain $\bm{X_C}$.}

2. When considering the coefficient of each covariate $j$, the weight $\hat{w}_j$ will be always less or equal to 1, therefore, $\sum_{i=1}^p\hat{w}_j^2$ in the second model will be less than $p$. From the theory in \textbf{Online Supplement S1 Appendix F}, the estimator in the second model is always a root-n/p-consistent estimator. 

Now we consider the possibility of adopting the tanh function as the penalty smoothing function, as it can map $[0, \infty)$ to $[0, 1)$. According to the theory in \textbf{Online Supplement S1 Appendix F}, it seems that the tanh function not only retains the advantages of the sigmoid function but also restricts the value of $E||\bm{\hat{\beta}(\lambda_2,\lambda_1)-\beta^*}||_2^2$ within the range of $(0, p)$ instead of $(0.5p, p)$, and provide a relatively more accurate elastic net estimator.

Unfortunately, the tanh function does not outperform the sigmoid function in the synthetic datasets regarding the preliminary two-stage framework and it faces the challenge of eliminating $\bm{X_S}$. Since the exposure model assigns weights of $\bm{X_S}$ to be close to 0, the tanh function does not provide enough penalty to these covariates. However, the sigmoid function sets the penalty factor to a non-zero value in the outcome model, thus reserving some penalization ability to filter $\bm{X_S}$. Additionally, note that the sigmoid function has weaker convexity than the tanh function, the sigmoid function will not separate covariates too much. In contrast, the tanh function imposes a higher penalty on $\bm{X_C}$ than the sigmoid function. Consequently, it potentially leads to an empty set of selected variables.

%%%%%%%%%%%%%%%Section 4.3%%%%%%%%%%%%%%%%%%
\subsection{Selecting $\gamma$} \label{sec:a_gamma}
%%%%%%%%%%%%%%%Section 4.3%%%%%%%%%%%%%%%%%%

In this subsection, we present guidance in selecting $\gamma$ in both Framework A and Framework B on the lasso-based structure to support our statement that Framework A requires less effort in tuning parameters.

%%%%%%%%%%%%%%%Section 4.3.1%%%%%%%%%%%%%%%%%%
\subsubsection{Selecting $\gamma$ in lasso-based method of Framework B}
%%%%%%%%%%%%%%%Section 4.3.1%%%%%%%%%%%%%%%%%%
In Framework B, the wAMD equation is the equation to be optimized. To meet the oracle property, first, it defines the gamma converge factor $\Gamma$, then chooses the best pair $(\lambda_n, \gamma)$ that minimizes wAMD. 

Now let's consider the boundary of $\gamma$ in Framework B. Since the oracle property requires two conditions: $\lambda_n/\sqrt{n} \rightarrow 0$ and $\lambda_n n^{\gamma/2-1} = n^{\Gamma} \rightarrow \infty$ (Note that in general cases $\lambda_n n^{(\gamma-1)/2-1}\rightarrow \infty$. However, \cite{shortreed2017outcome} points out that in causal inference, variables that predict exposure should be excluded, thus a stronger condition is required). The second condition gives $\Gamma \geq 0$, and the first condition yields $\Gamma+1-\gamma/2<0.5$. Thus, $\gamma>2\Gamma+1$ and $\Gamma \geq 0$. However, there is no upper bound for $\gamma$, which makes tuning $\gamma$ a challenging process because minimizing wAMD is a nonlinear optimization problem with respect to $\gamma$.

%%%%%%%%%%%%%%%Section 4.3.2%%%%%%%%%%%%%%%%%%
\subsubsection{Selecting $\gamma$ in lasso-based method of Framework A}
%%%%%%%%%%%%%%%Section 4.3.2%%%%%%%%%%%%%%%%%%
Compared to Framework B, $\gamma$ in Framework A is less sensitive because the outcome model can adjust the prior knowledge by itself. $\gamma$ in Framework A is more interpretable as it functions as the adjustment factor of convexity (see \textbf{Online Supplement S1 Appendix J}). Our experiments find that if we use the sigmoid function as the penalty smoothing function, $\gamma=1$ should be fine. Applying a large $\gamma$ further separates the penalty of $\bm{X_C}$ and $\bm{X_S}$, decreasing the selection ability of $\bm{X_C}\cup\bm{X_T}$ and increasing the selection ability of $\bm{X_P}\cup\bm{X_S}$. Conversely, applying a smaller value of $\gamma$ increases the selection ability of $\bm{X_C}\cup\bm{X_T}$ while decreasing the selection ability of $\bm{X_P}\cup\bm{X_S}$.

%%%%%%%%%%%%%%%Section 4.4%%%%%%%%%%%%%%%%%%
\subsection{Limitation of the preliminary two-stage framework} \label{sec:limit}
%%%%%%%%%%%%%%%Section 4.4%%%%%%%%%%%%%%%%%%
Although efficient in selecting variables proposed in Proposition \ref{prop:ideal_var_sel}, unfortunately, we cannot guarantee that the preliminary models satisfy the oracle property. This is because the selected $\lambda$ in Section \ref{sec:lambda_Sel} is designed to minimize the cross-validation error of the outcome model, but the model not necessarily meet the two conditions required for the oracle property (i.e., $\lambda_1/\sqrt{n} \rightarrow 0$ and $\lambda_1n^{(\gamma-1)/2-1} \rightarrow \infty$). One might attempt to search for the optimal $\lambda$ that minimizes the cross-validation error while satisfying the constraints given by the two conditions of the oracle property. However, this approach would be challenging and time-consuming, as $\lambda$ does not have an upper bound.

Luckily, there is an alternative approach that both meets the oracle property and requires less effort in tuning hyperparameters. According to the mathematical proof presented in \textbf{Online Supplement S1 Appendix F}, any non-negative parameters $\lambda_1$ and $\lambda_2$ will make estimations of coefficients from the preliminary model a root-$n/p$-consistent estimator. Such an estimator is considered suitable for the adaptive elastic net model \citep{zou2009adaptive}. This approach compromises on time complexity to achieve the oracle property by fitting an additional adaptive elastic net model in the outcome model, utilizing the estimation from Eq.(\ref{model:two-stage-outcome-model}). 

%%%%%%%%%%%%%%%Section 5%%%%%%%%%%%%%%%%%%
\section{Enhanced Three-Stage Variable Selection Framework for Causal Inference} \label{model:three-stage-framework}
%%%%%%%%%%%%%%%Section 5%%%%%%%%%%%%%%%%%%

In this section, we propose the three-stage framework. This framework includes one exposure model and two outcome models. The three-stage framework adds an additional adaptive elastic net model in the outcome model in order to keep the oracle property while reducing the burden of tuning hyperparameters. First, it fits the exposure model by SVM or LR estimator. Then, it inputs the prior knowledge to the first model of outcome models, which is the elastic net model. At last, it inputs the coefficient of each covariate in the elastic net model to the second model of outcome models.

\texttt{exposure model: 
    \begin{equation} \label{model:three-model-exp}
        \hat{\bm{\beta}} = \hat{\bm{\beta}}(SVM) ~ or ~ \hat{\bm{\beta}}(LR) 
    \end{equation}
} 

\texttt{outcome models: 
    \begin{equation} \label{model:three-model-out-el}
        \bm{\hat{\theta}} = argmin_{\bm{\theta}}||\bm{Y}-\bm{X\theta}||_2^2 + \lambda_2^{(1)}||\bm{\theta}||_2^2+ \lambda_1^{(1)} \sum_{j=1}^{p}{\hat{w}_j|\theta_j|}
    \end{equation}
    \begin{equation} \label{model:three-model-out-adel}
        \bm{\hat{\Theta}} = (1 + \frac{\lambda_2^{(2)}}{n})
        \bigg\{ argmin_{\bm{\Theta}}||\bm{Y}-\bm{X\Theta}||_2^2 + \lambda_2^{(2)} ||\bm{\Theta}||_2^2+ \lambda_1^{(2)} \sum_{j=1}^{p}{\hat{\Psi}_j|\Theta_j|} \bigg\}
    \end{equation}
}

Here $\hat{w}_j=\big\{ \frac{1}{1+exp(-|\hat{\beta}_j|)} \big\}^{\gamma_1} $  or $\hat{w}_j=\big\{ \frac{exp(|\hat{\beta}_j|)-exp(-|\hat{\beta}_j|)}{exp(|\hat{\beta}_j|)+exp(-|\hat{\beta}_j|)} \big\}^{\gamma_1}$, the former function is known as sigmoid function, and the latter function is known as tanh function. $\hat{\Psi}_j=(\frac{1}{|\hat{\theta}_j|})^{\gamma_2}$ and $\gamma_2$ are hyperparameters and should be selected via the assumptions outlined in \textbf{Online Supplement S1 Appendix E}.

For simplicity in notation, we add the prefix Enh in naming the models, which refers to the enhanced three-stage framework. For example, if the three-stage framework uses logistic regression as the estimator in the exposure model and uses the tanh function as the penalty smoothing function, then we name the model \texttt{Enh-ELRT}.

For the rest of this section, we discuss the advantages of the proposed three-stage framework over the preliminary two-stage model. Then we analyze the proper settings and selections of the proposed framework.

%%%%%%%%%%%%%%%Section 5.1%%%%%%%%%%%%%%%%%%
\subsection{Improvement compared to the two-stage framework} 
%%%%%%%%%%%%%%%Section 5.1%%%%%%%%%%%%%%%%%%

The three-stage framework has some advantages over the preliminary two-stage framework. As presented in Section \ref{sec:limit}, the preliminary model faces the challenge of simultaneously meeting the oracle property and tuning the optimal $\lambda$. The three-stage framework mitigates the sensitivity of selecting $\lambda$ by handing it over to the second outcome model (the adaptive elastic net model). Consequently, it makes it easier and interpretable for tuning the hyperparameter $\gamma_2$ (we go depth in section \ref{sec:choose_gamma_2}). 

It's worth noting that in the three-stage framework, we utilize the outcome label twice in the outcome model. Utilizing the outcome label further, however, will not improve the results, as the oracle property can be certainly satisfied with our three-stage framework. 

%%%%%%%%%%%%%%%Section 5.1%%%%%%%%%%%%%%%%%%
\subsection{Choosing penalty smoothing function and selecting $\gamma_1$} \label{sec:choose_gamma_1}
%%%%%%%%%%%%%%%Section 5.1%%%%%%%%%%%%%%%%%%

In the three-stage framework, unlike the preliminary framework, either the sigmoid function or the tanh function can be used as the penalty smoothing function. If we use the tanh function as the penalty smoothing function, even though model (\ref{model:three-model-out-el}) assigns a non-zero coefficient to $\bm{X_S}$, model (\ref{model:three-model-out-adel}) can still eliminate $\bm{X_S}$ because high penalties are assigned to these covariates.

For the selection of $\gamma_1$, if we use the sigmoid function, $\gamma_1=1$ should suffice. However, if we use the tanh function, based on our experiments, we recommend setting $\gamma_1$ to be a value smaller than 1. This is because the tanh function exhibits relatively strong convexity and may impose excessive penalties on $\bm{X_C}$, potentially resulting in the exclusion of $\bm{X_C}$ during the feature selection process. In our experiments, we set $\gamma_1=0.5$.

%%%%%%%%%%%%%%%Section 5.2%%%%%%%%%%%%%%%%%%
\subsection{Selecting $\gamma_2$} \label{sec:choose_gamma_2}
%%%%%%%%%%%%%%%Section 5.2%%%%%%%%%%%%%%%%%%

Choosing $\gamma_2$ in our proposed model is a relatively easy task. According to the assumptions in \textbf{Online Supplement S1 Appendix E}, $\gamma_2 \geq \lceil \frac{2v}{1-v} \rceil+1$, where $v=\lim_{n\rightarrow{\infty}}\frac{log(p)}{log(n)}$, and $0\leq v \leq 1$. Thus, $\gamma_2 \geq 1$. Therefore, it is safe to set $\gamma_2=1$. We suggest setting $\gamma_2=1$ because as $\gamma_2$ increases, the selection ability of $\bm{X_C}$ will decrease. However, our experiments also indicate that when $\gamma_2$ is set to values larger than 1, the subset of selected variables remains unchanged even when the covariates are highly correlated. This observation suggests that this framework is robust and not sensitive to the choice of $\gamma_2$.

%%%%%%%%%%%%%%%Section 5.3%%%%%%%%%%%%%%%%%%
\subsection{Properties of the proposed three-stage framework} \label{sec:prop_prop_mdl}
%%%%%%%%%%%%%%%Section 5.3%%%%%%%%%%%%%%%%%%

In our proposed model, regardless of the choice of penalty smoothing function (sigmoid function or tanh function) and the exposure model (SVM or LR estimator), the estimator (\ref{model:three-model-out-el}) is always a root-n/p consistent estimator. According to the \textbf{Online Supplement S1 Appendix F}, the oracle property of the proposed three-stage framework is well-defined as follows:

\begin{theorem}[Oracle Property of Proposed Model] \label{the:3-model}
Given data $(\bm{Y}, \bm{X}, \bm{T})$, let $\hat{\bm{w}} = (\hat{w}_1, \hat{w}_2 ..., \hat{w}_p)$ be a vector whose components are all non-negative, then under six assumptions listed in the \textbf{Online Supplement S1 Appendix E} \citep{zou2009adaptive}, the adaptive elastic net reaches the oracle property: \\
1. Consistency in variable selection: $Pr(\{j|\hat{\beta}(AdaEnet)_j\neq0\}= \mathcal{A})\rightarrow{1}$ \\
2. Asymptotic normality: $\bm{u^T}\frac{\bm{I}+\lambda_2\bm{\Sigma_{11}^{-1}}}{1+\lambda_2/n}\bm{\Sigma_{11}^{1/2}}(\hat{\beta}(AdaEnet)_\mathcal{A}-\beta_{\mathcal{A}}^*) \rightarrow{_dN(0,\sigma^2)}$
\end{theorem}

Here $\bm{u^T}$ is an imposed vector which satisfies $\bm{u^T}\bm{u}=1$, $\bm{\Sigma_{11}}$ is known as the Fisher Information Matrix and is the upper left block of $\bm{X^T}\bm{X}$\remove[tianyu]{with dimension $p_0*p_0$}. The mathematical proof for the oracle property \change[tianyu]{can follow}{follows} \textbf{Online Supplement S1 Appendix F}. 

We now discuss the selection ability of different settings. Two pairs of general settings may impact the final output: the choice of exposure model (SVM or logistic regression), and the choice of the penalty smoothing function (tanh or sigmoid function). 

Before we discuss the two settings, it is worth highlighting that a better $\bm{\hat{\Theta}}$ must be estimated from a better $\hat{\bm{\theta}}$ always holds true, as the structure of the outcome models always keeps the same. A better $\hat{\bm{\theta}}$ implies that this estimator is more robust and strongly suggests the subset of variables listed in Proposition \ref{prop:ideal_var_sel}. Therefore, the analysis of choosing a better $\bm{\hat{\Theta}}$ is equivalent to choosing a better $\hat{\bm{\theta}}$. In the experiment, a direct method to determine the superior one is to compare the coefficient output from the outcome model, as the estimated coefficients serve as an indicator of variable selection, we present related experiment results in \textbf{Online Supplement S1 Appendix D}. Intuitively, for the choice of exposure model, a logistic regression estimator tends to assign extreme weights to $\bm{X_T}$ and $\bm{X_C}$ and assign small weights to $\bm{X_P}$ and $\bm{X_S}$. Since for the first model of the outcome model $\hat{w_j} \propto |\hat{\beta}_j|$, compared to SVM, it has a stronger ability to eliminate $\bm{X_T}$ and choose $\bm{X_P}$, but a weaker ability to choose $\bm{X_C}$. For the penalty smoothing function, since the tanh function has stronger convexity, similar to logistic regression, it tends to assign very small or large penalties. Therefore, compared to the sigmoid function, it has a stronger ability to eliminate $\bm{X_T}$ and choose $\bm{X_P}$, but a weaker ability to choose $\bm{X_C}$. 

\begin{proposition} [Selection Ability of Proposed Model] \label{prop:sel_ability}

    Under the proposed three-stage framework, applying a logistic regression estimator and tanh function will increase the ability to eliminate $\bm{X_T}$ and choose $\bm{X_P}$, but lack the ability to choose $\bm{X_C}$. On the other hand, applying an SVM estimator and sigmoid function will increase the ability to choose $\bm{X_C}$, but lack the ability to eliminate $\bm{X_T}$ and choose $\bm{X_P}$.
    
\end{proposition}

%%%%%%%%%%%%%%%Section 6%%%%%%%%%%%%%%%%%%
\section{Result Analysis for Synthetic Datasets} \label{sec:res_ana}
%%%%%%%%%%%%%%%Section 6%%%%%%%%%%%%%%%%%%

This section discusses our experimental results on synthetic datasets. All the experiments are run on the 2.4 GHz Quad-Core Intel Core i5 CPU using R.

\subsection{Benchmark on Causal Feature Selection Models}

To evaluate the ability of variable selection discussed in Section \ref{lemma: ubest}, we utilized one of the most widely used state-of-the-art synthetic datasets \citep{shortreed2017outcome, lu2020feature, balde2022reader}, which is generated based on the known confounders ($\bm{X_C}$), pure treatment predictors ($\bm{X_T}$), pure outcome predictors ($\bm{X_P}$) and noise variables ($\bm{X_S}$). 

We perform our experiments in the following 4 scenarios. $\bm{\theta}$ refers to the regression coefficients in the outcome model and $\bm{\beta}$ refers to the regression coefficients in the treatment model. The experiments are conducted with varying levels of correlation among covariates $\rho = \{0, 0.25, 0.5, 0.75\}$, a fixed number of covariates $p = 20$, and different numbers of generated cases $N = \{200, 500, 1000\}$. For each combination ($N$, $\rho$) in each scenario, the experiment is conducted 30 times with seeds ranging from 1 to 30 for reproducibility. Hence, the total number of experiments conducted is $4 \times 3 \times 4 \times 30 = 1,440$ runs.

\textbf{Scenario 1}. $\bm{\theta}=(0.6,0.6,0.6,0.6,0,0,0,...,0)$ and $\bm{\beta} = (1,1,0,0,1,1,0,...,0)$

\textbf{Scenario 2}. $\bm{\theta}=(0.6,0.6,0.6,0.6,0,0,0,...,0)$ and $\bm{\beta} = (0.4,0.4,0,0,1,1,0,...,0)$

\textbf{Scenario 3}. $\bm{\theta}=(0.2,0.2,0.6,0.6,0,0,0,...,0)$ and $\bm{\beta} = (1,1,0,0,1,1,0,...,0)$

\textbf{Scenario 4}. $\bm{\theta}=(0.6,0.6,0.6,0.6,0,0,0,...,0)$ and $\bm{\beta} = (1,1,0,0,1.8,1.8,0,...,0)$

Recall the definition of $\bm{\theta}$ and $\bm{\beta}$, in scenarios 1 to 4, $\bm{X_P}=\{3,4\}$, $\bm{X_T}=\{5,6\}$, $\bm{X_C}=\{1,2\}$, the variables to be chosen should be $\bm{X_P}\cup\bm{X_C}=\{1,2,3,4\}$.

%%%%%%%%%%%%%%%Section 6.1%%%%%%%%%%%%%%%%%%
\subsubsection{Benchmark Models} \label{sec:methods_of_imp}
%%%%%%%%%%%%%%%Section 6.1%%%%%%%%%%%%%%%%%%

We utilize the following four models derived from the proposed three-stage framework to evaluate their performance (i) \texttt{Enh-ELRT}, with logistic regression estimator and tanh penalty smoothing function.; (ii) \texttt{Enh-ELRS}, with logistic regression estimator and sigmoid penalty smoothing function.; (iii) Enhanced three-stage framework \texttt{(Enh-ESVMT)}, with SVM estimator and the tanh penalty smoothing function. (iv) \texttt{(Enh-ESVMS)}, with SVM estimator and the sigmoid penalty smoothing function..

We benchmarked our proposed feature selection framework with the following state-of-the-art lasso-based framework for our performance evaluation (v) Adaptive Lasso \texttt{(OAL)} Method \citep{shortreed2017outcome} with wAMD method, utilizing R package \texttt{lqa(v 1.0.3)}; (vi) Outcome Adaptive Elastic Net \texttt{(OAENet)} Method  \citep{islam2021feature} with wAMD method using R package \texttt{glmnet}. We also use the following state-of-the-art Bayesian theory-based framework for evaluation (vii) \texttt{BACR} method from \citep{wang2012bayesian} using R package \textit{bacr}; (Viii) \texttt{BCEE} method from \citep{talbot2015bayesian} using R package \textit{BCEE}. Additionally, we use the following state-of-the-art causal-tree-based models for evaluation (ix) Boruta to identify treatment predictors (\texttt{Boruta-T}) and (x) Boruta to identify outcome predictors (\texttt{Boruta-Y}) using R package \textit{Boruta}, (xi) DeLuna, Waernbaum and Richardson (\texttt{DWR}) \citep{de2011covariate} using R package \texttt{CovSel}, (xii) Collaborative Targeted Maximum Likelihood Estimation (\texttt{CTMLE}) \citep{ju2019scalable} using R package \texttt{ctmle(v 0.1.2)}. 

%%%%%%%%%%%%%%%Section 6.1.2%%%%%%%%%%%%%%%%%%
\subsubsection{Selection Bias to the True ATT estimation} \label{res:ATT_est}
%%%%%%%%%%%%%%%Section 6.1.2%%%%%%%%%%%%%%%%%%

The experiment is designed as follows: each model first identifies the selected features. Then, the one-to-one nearest matching technique is employed based on these selected variables to filter the matched groups of treatment and control. Then we estimate ATT based on the regression model and record the absolute difference between the ATT estimation and the true ATT values as selection bias, given by the regression model which includes all the predefined $\bm{X_P}\cup\bm{X_C}$ (also known as the target model (\texttt{TM})
). 

\begin{figure}[htbp]
    \centering
    \subfloat[Scenario 1 selection bias for $N=200,p=20$]{%
        \includegraphics[width=0.44\textwidth]{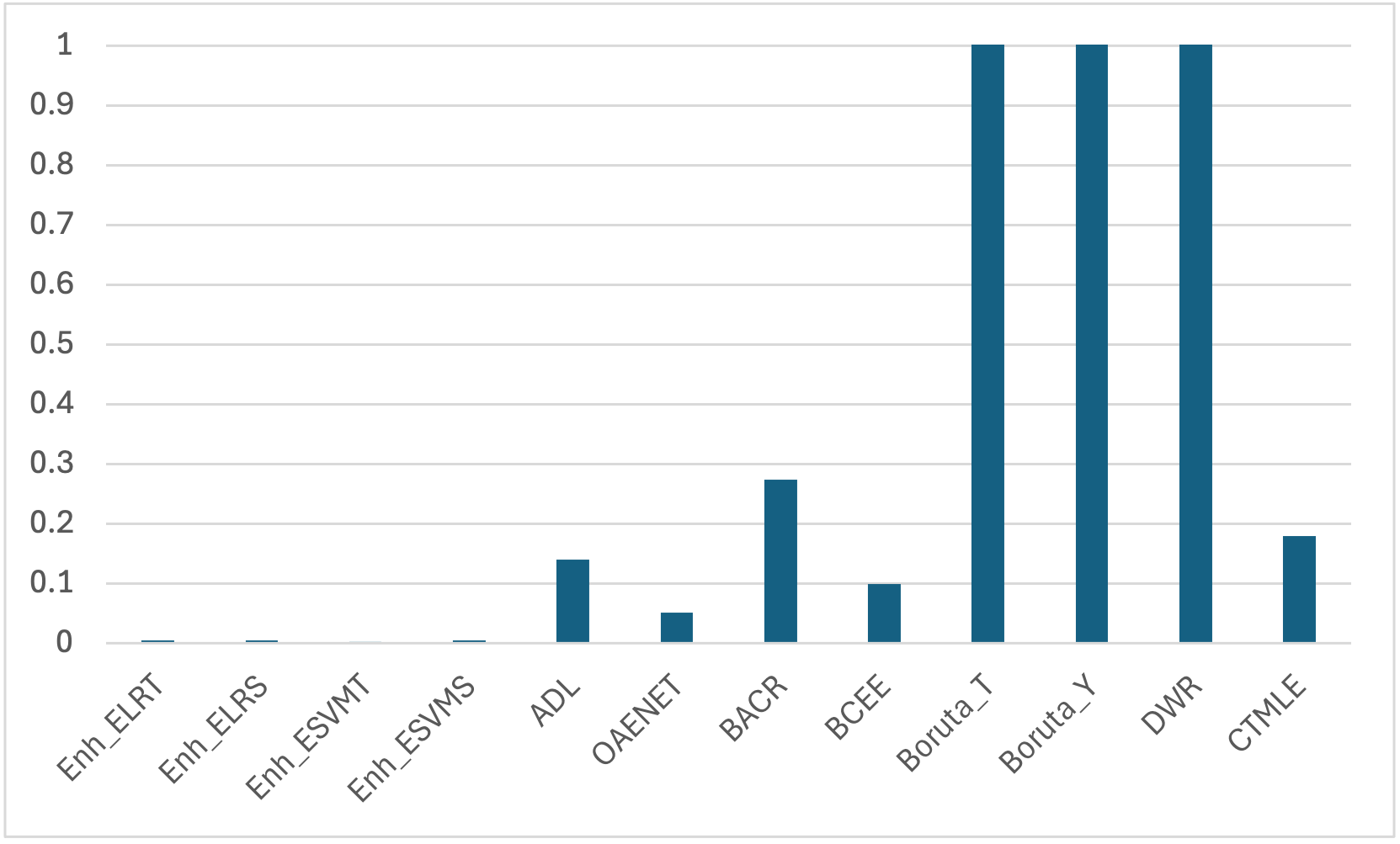}%
    }%
    \hfill
    \subfloat[Scenario 2 selection bias for $N=200,p=20$]{%
        \includegraphics[width=0.44\textwidth]{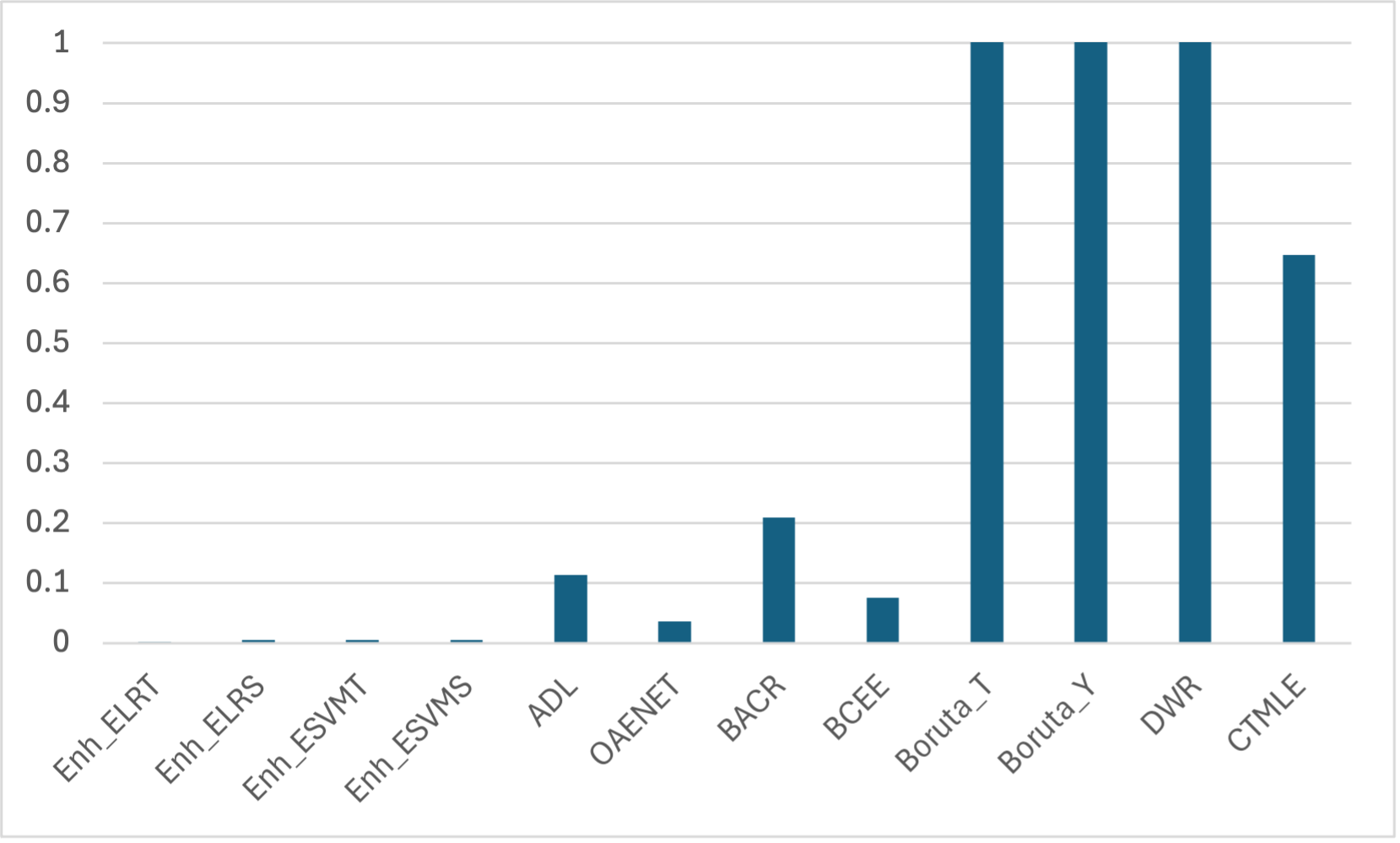}%
    }%
    \hfill
    \subfloat[Scenario 3 selection bias for $N=200,p=20$]{%
        \includegraphics[width=0.44\textwidth]{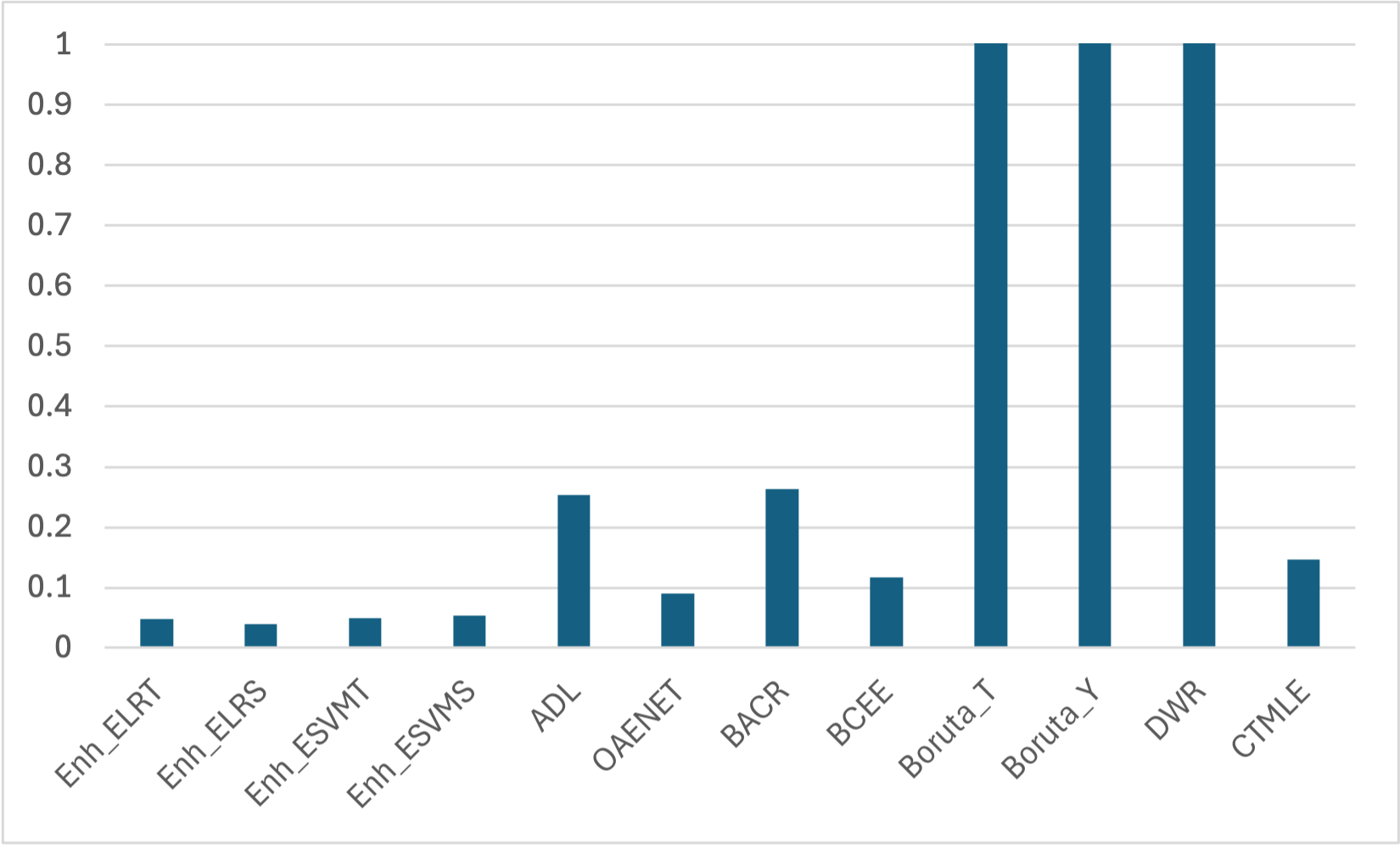}%
    }%
    \hfill
    \subfloat[Scenario 4 selection bias for $N=200,p=20$]{%
        \includegraphics[width=0.44\textwidth]{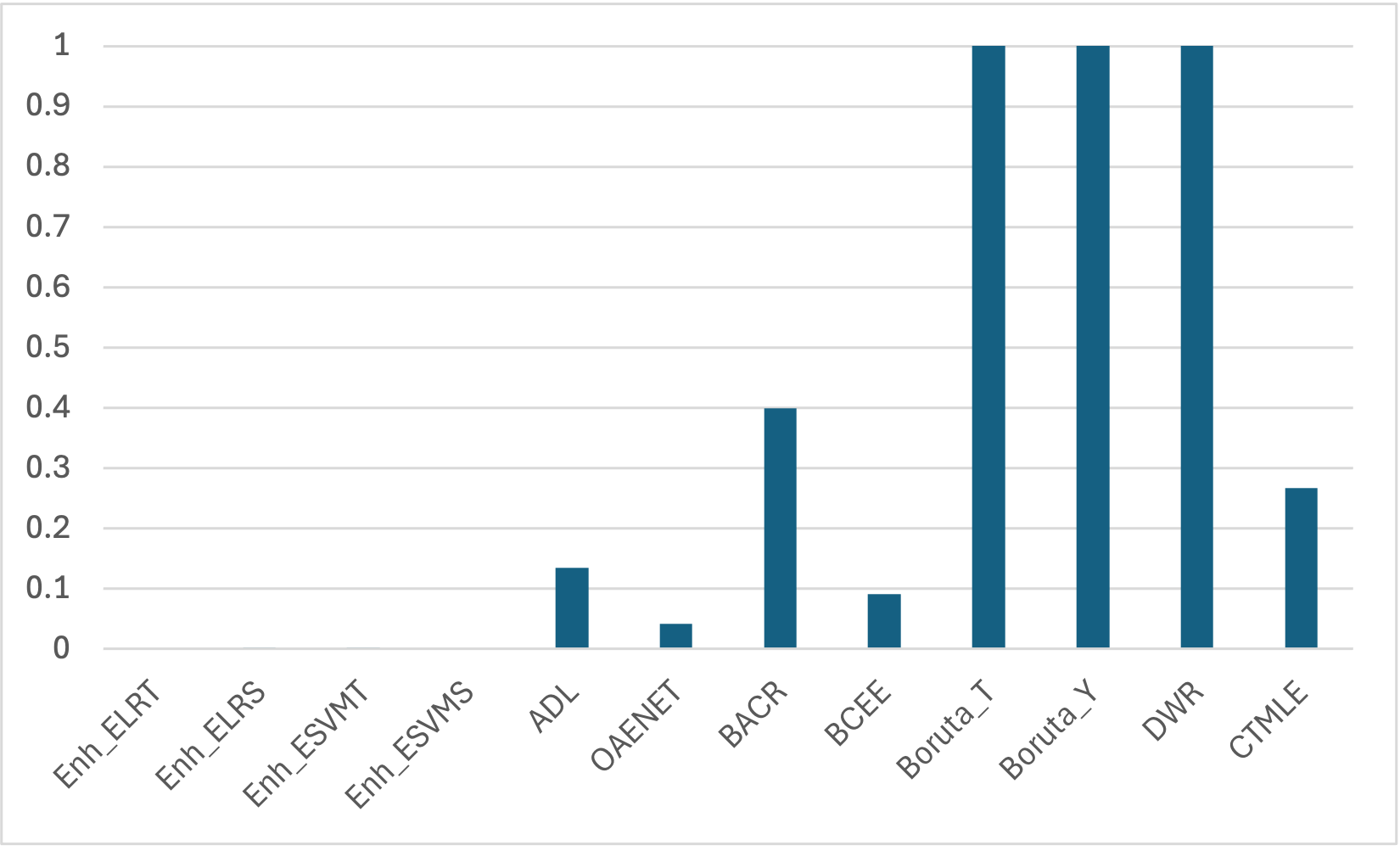}%
    }%
    \caption{The selection bias to the ATT estimation for each model.}
    \label{fig:atts}
\end{figure}

We present the selection bias to the ATT estimation for each model with $N=200$. The ideal framework is supposed to provide 0 bias. Figure \ref{fig:atts} displays the ATT estimation across $\rho=\{0, 0.25, 0.5, 0.75\}$. Results for $N=500$ and $N=1000$ results are available in \textbf{Online Supplement S1 Appendix H.2.1}.

Figure \ref{fig:atts}, along with the figures recording bias to the ATT estimation provided in \textbf{Online Supplement S1 Appendix H.2.1}, demonstrates the significantly remarkable consistency in 
achieving significantly low bias compared to other state-of-the-art methods across all scenarios. To further demonstrate the superiority of our proposed feature selection framework, we conducted hypothesis tests on the absolute selection bias equal to 0. The results showed that our framework consistently selects variables that produce ignorable selection bias, whereas the state-of-the-art methods including BCEE and OAENet do not have such property. Detailed results are provided in \textbf{Online Supplement S1 Appendix G}.

\subsubsection{Probability of each variable being selected} \label{res:var_sel}
%%%%%%%%%%%%%%%Section 6.4%%%%%%%%%%%%%%%%%%

We present the probability of each variable being selected by our proposed models and benchmark models for $N=200$. Results for $N=500$ and $N=1000$ are available in \textbf{Online Supplement S1 Appendix H.2.2}.

\begin{figure}[htbp] 
    \centering
    \subfloat[Scenario 1 $N=200,p=20$]{%
        \resizebox{0.44\textwidth}{2.0in}{
        \includegraphics[width=\textwidth]{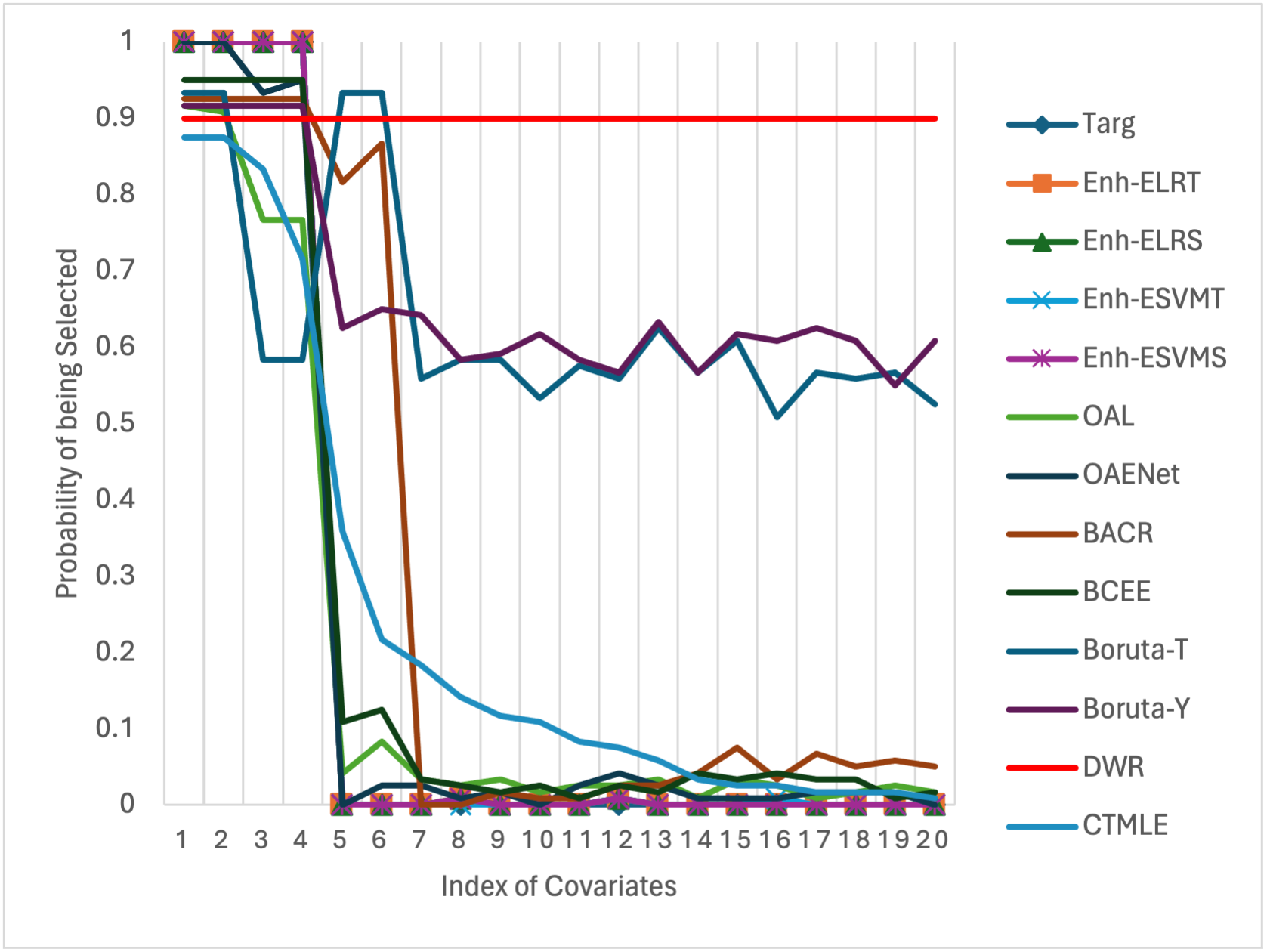}%
        }
    }%
    \hfill
    \subfloat[Scenario 2 $N=200,p=20$]{%
        \resizebox{0.44\textwidth}{2.0in}{
        \includegraphics[width=\textwidth]{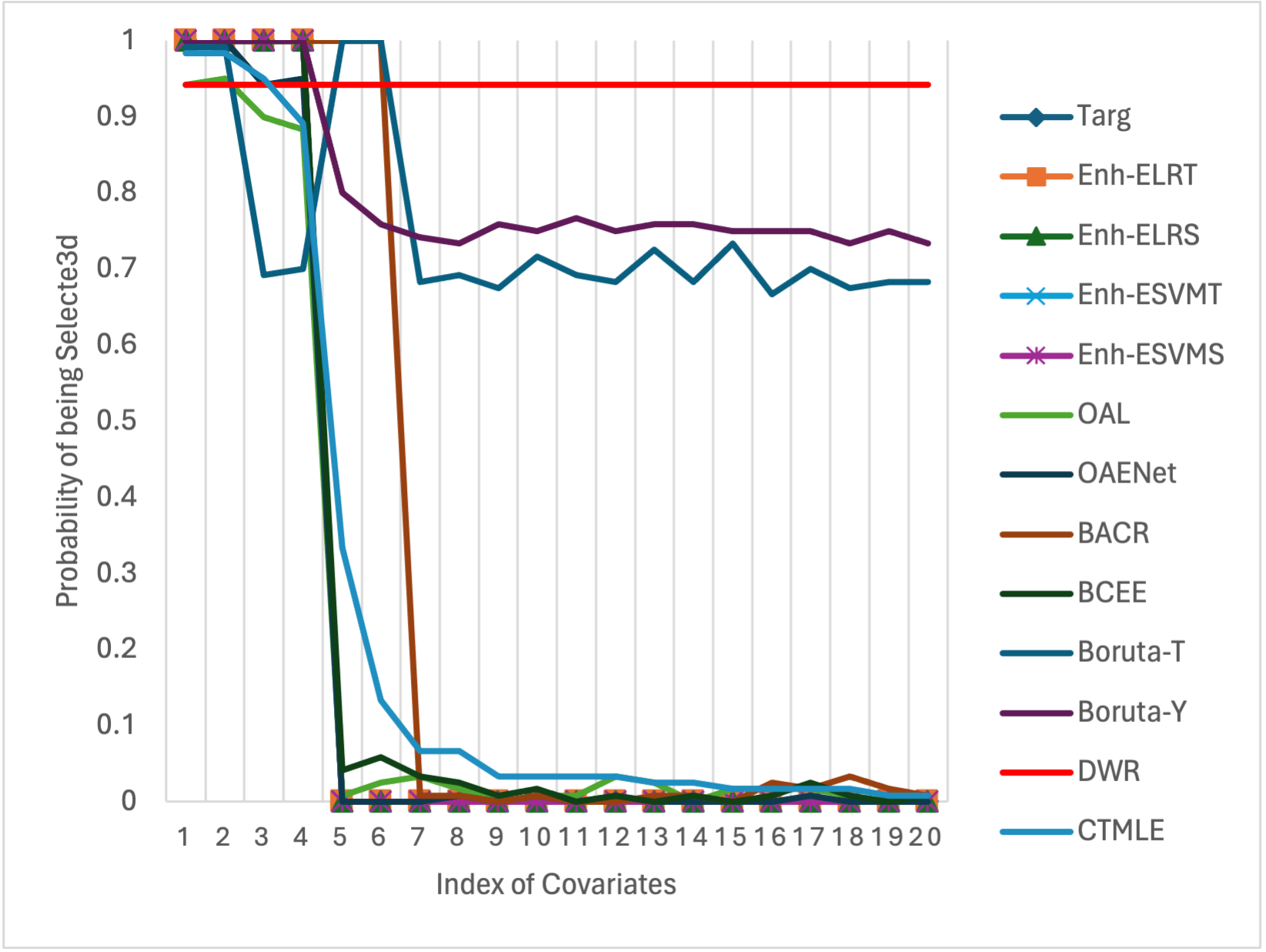}%
        }
    }%
    \hfill
    \subfloat[Scenario 3 $N=200,p=20$]{%
        \resizebox{0.44\textwidth}{2.0in}{
        \includegraphics[width=\textwidth]{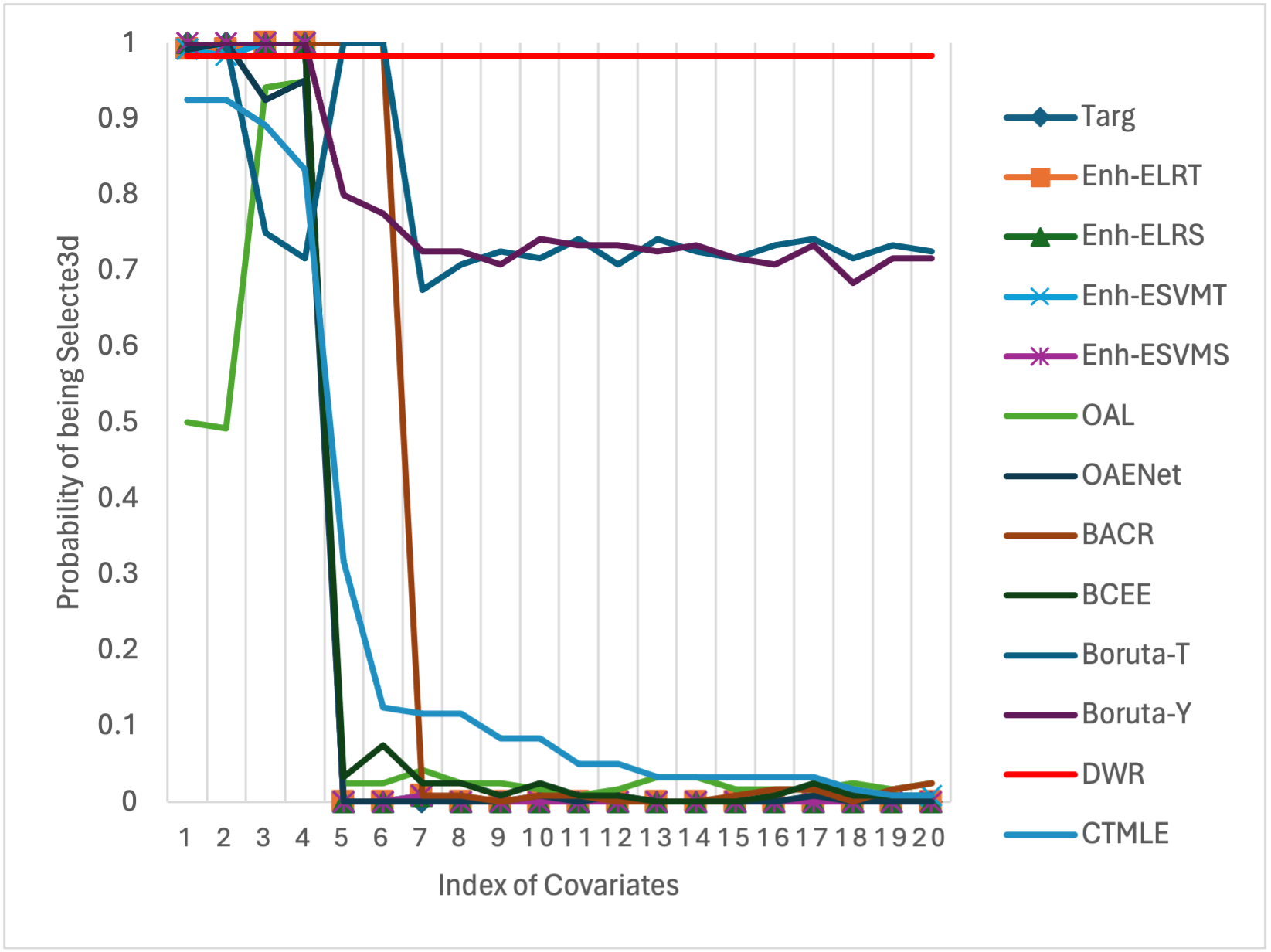}%
        }
    }%
    \hfill
    \subfloat[Scenario 4 $N=200,p=20$]{%
        \resizebox{0.44\textwidth}{2.0in}{
        \includegraphics[width=\textwidth]{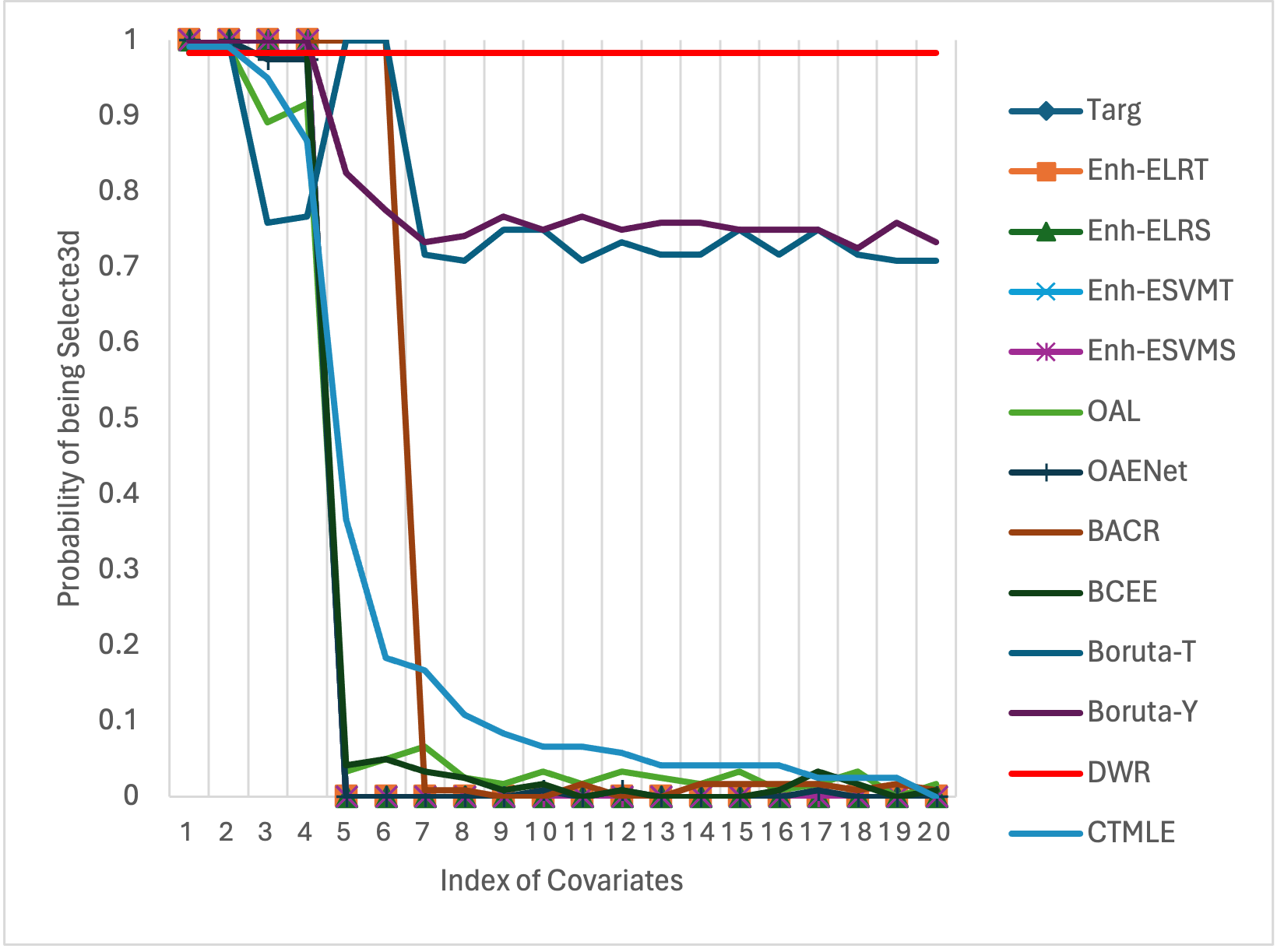}%
        }
    }%
    \caption{Variable selection across all the scenarios. The x-axis refers to the index of the covariates, while the y-axis indicates the probability of each variable being selected.}
    \label{fig:vars}
\end{figure}

In each figure in Figure \ref{fig:vars}, the horizontal axis represents the index of each covariate, while the vertical axis indicates the probability of each covariate being selected by each model. We highlight the target model and our proposed three-stage framework models with markers and keep the lines of other benchmark models without markers. 

Based on the results presented in Figure \ref{fig:vars}, along with the plots in \textbf{Online Supplement Appendix H.2.2}, our proposed framework demonstrates superior performance by consistently chooses $\bm{X_P} \cup \bm{X_C}$ with probability near 1 and choose $\bm{X_T} \cup \bm{X_S}$ with probability near 0. Noticing that the four proposed three-stage frameworks have similar performance, further experiments are conducted in \textbf{Online Supplement Appendix H.2.1} to evaluate the logistic regression-based three-stage models and SVM-based three-stage models. Results show that logistic regression-based three-stage models perform better with small datasets ($N$ is small) and high-correlated structure datasets ($\rho$ is high), while SVM-based three-stage models perform better in the rest of the cases. However, when the size of the dataset increases, SVM-based three-stage models seem to overcome the shortage of correlation and always pass smaller penalties of $\bm{X_C}$ to the outcome model. Thus, they have a higher ability to choose $\bm{X_C}$ than logistic regression-based models. This observation agrees with the argument made by Proposition \ref{prop:sel_ability}.  

Since the results demonstrate the effectiveness of our proposed framework in selecting variables suggested by Proposition \ref{prop:ideal_var_sel}, we conclude that our proposed framework can effectively reduce bias and variance when estimating ATT.

%%%%%%%%%%%%%%%Section 6.5%%%%%%%%%%%%%%%%%%
\subsubsection{Computational time}
%%%%%%%%%%%%%%%Section 6.5%%%%%%%%%%%%%%%%%%

We present the computational time for each method in Table \ref{tbl:compute_time} for $N=200$ records, by averaging the CPU time across all the scenarios. According to Table \ref{tbl:compute_time}, OAL and OAENet exhibit faster computation times compared to other algorithms. Our proposed models, however, exhibit superior performance in reducing bias and variance in estimating ATT, while maintaining comparable computational efficiency.

\begin{table}[htbp]
    \centering
    \caption{Average CPU time to output the subset of variables for 200 observations (seconds)}
    \resizebox{\textwidth}{0.45in}{
    \begin{tabular}{||c|c||c|c||c|c||c|c||}
        \hline
        \textbf{Model} & \textbf{CPU Time} & \textbf{Model} & \textbf{CPU Time} & \textbf{Model} & \textbf{CPU Time} & \textbf{Model} & \textbf{CPU Time} \\
        \hline
        Enh\_ELRT   & 1.396 & Enh\_ELRS  & 1.363  & Enh\_ESVMT & 1.456 & Enh\_ESVMS & 1.437 \\
        \hline
        OAL    & 0.126  & OAENet & 0.394 & BACR   & 1.881 & BCEE   & 8.475 \\
        \hline
        Boruta\_T & 1.476 & Boruta\_Y & 2.705 & DWR    & 4.444 & CTMLE  & 2.652 \\ \hline
    \end{tabular}
    }
    
    \label{tbl:compute_time}
\end{table}

\begin{figure}[htbp] 
    \centering
    \subfloat[Bias for Scenario 1 $N=1000,p=20$]{%
        \resizebox{0.49\textwidth}{2.0in}{
            \includegraphics[width=\textwidth]{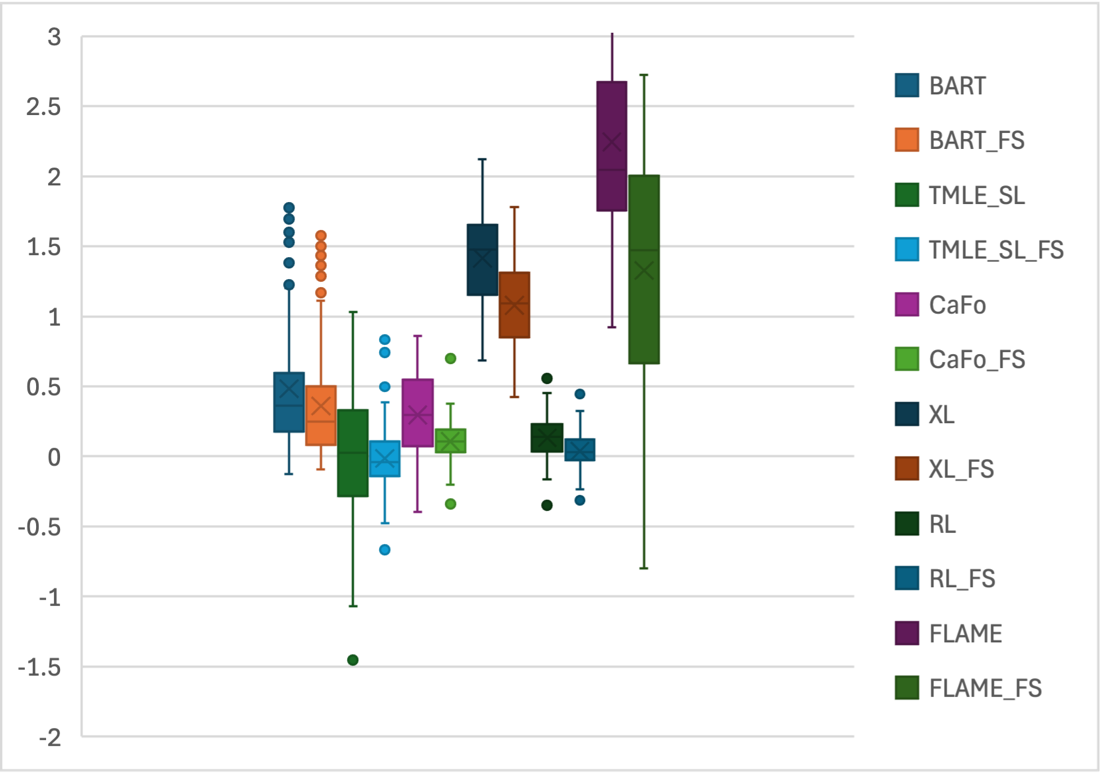}%
        }
    }%
    \hfill
    \subfloat[Bias for Scenario 2 $N=1000,p=20$]{%
        \resizebox{0.49\textwidth}{2.0in}{
        \includegraphics[width=\textwidth]{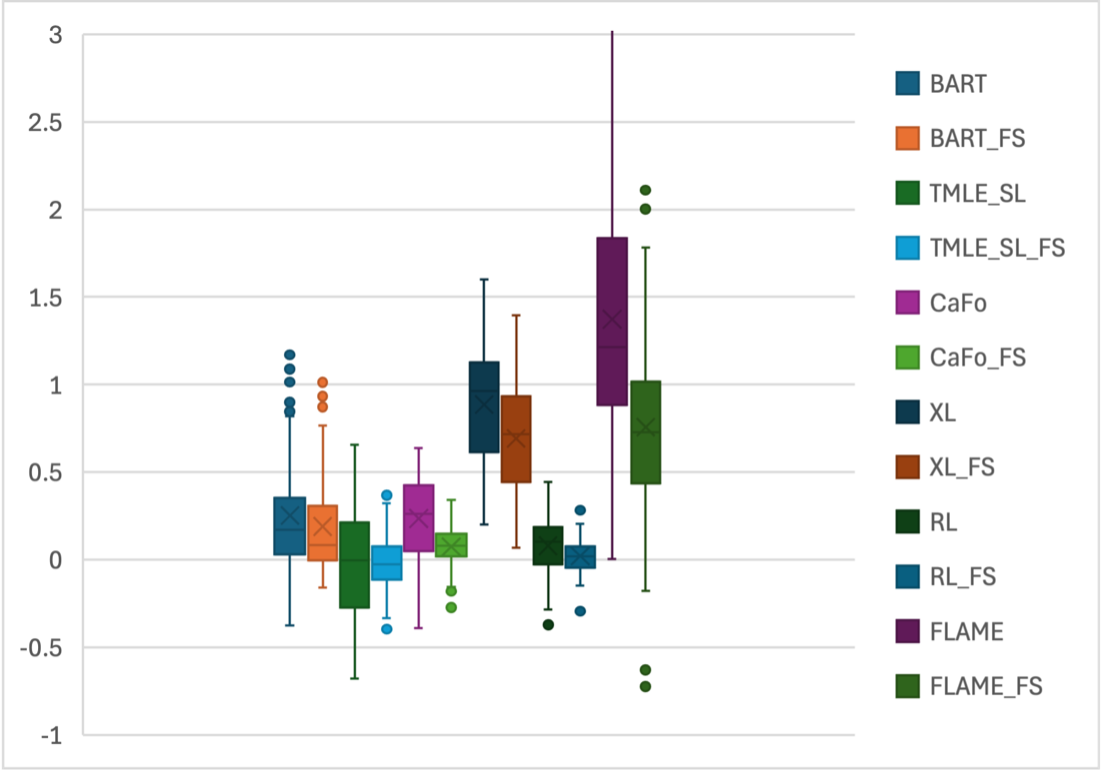}%
        }
    }%
    \hfill
    \subfloat[Bias for Scenario 3 $N=1000,p=20$]{%
        \resizebox{0.49\textwidth}{2.0in}{
        \includegraphics[width=\textwidth]{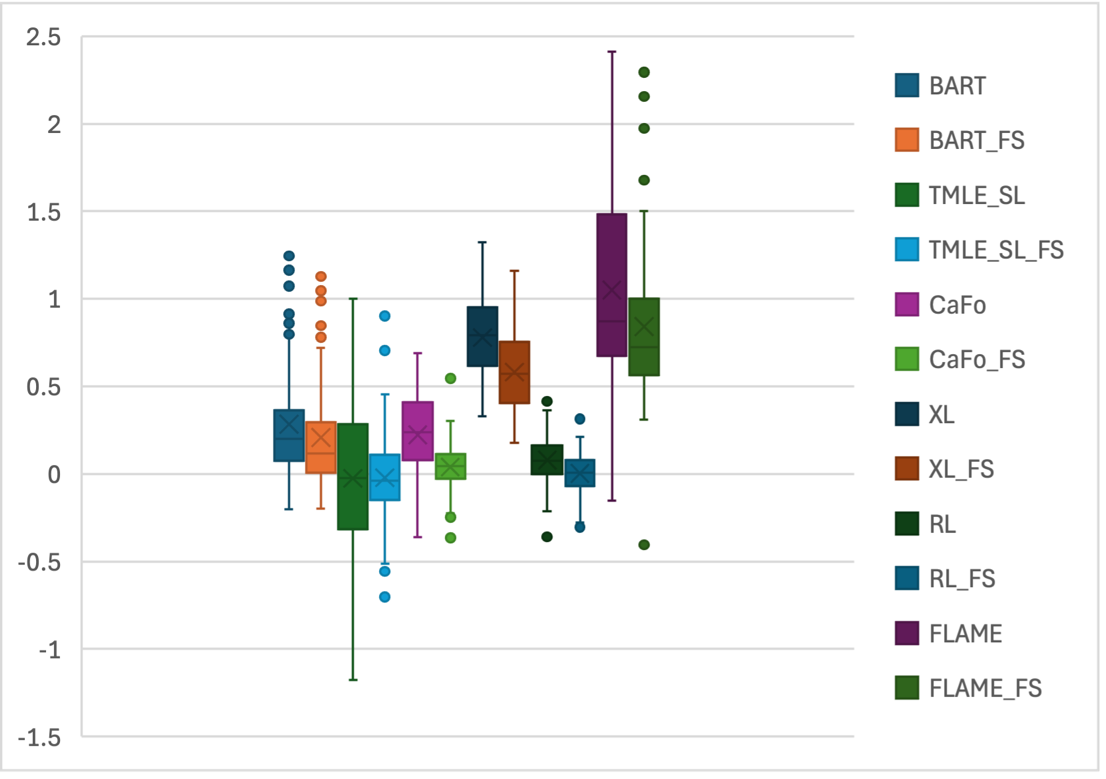}%
        }
    }%
    \hfill
    \subfloat[Bias for Scenario 4 $N=1000,p=20$]{%
        \resizebox{0.49\textwidth}{2.0in}{
        \includegraphics[width=\textwidth]{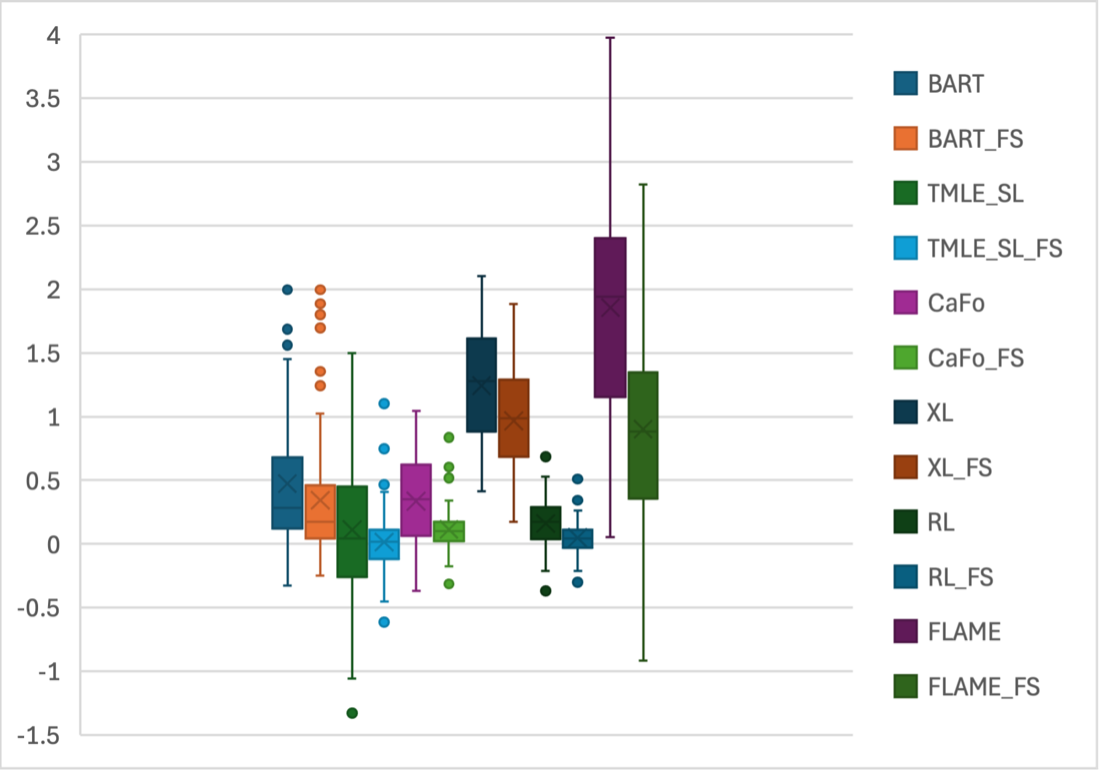}%
        }
    }%
    \caption{Bias of ATT estimation for causal machine learning models for each scenario.}
    \label{fig:TPM_bias}
\end{figure}

\subsection{Benchmark on Causal Machine Learning Models}

This subsection displays experimental results by combining the proposed feature selection framework with the popular causal inference machine learning methods. 

We used the following causal inference algorithms: BART, TMLE-SL, Causal Forest, X-Learner, R-Learner, and FLAME \citep{wang2021flame}. When adding “$\_$FS”, it means the method incorporates our proposed feature selection algorithm to select features first and then apply the causal inference machine learning algorithms through all the scenarios. For each scenario, we utilized 1,000 randomly generated units, 20 variables, with covariance among variables $\rho$ ranging from \{0, 0.25, 0.5, 0.75\}. Each combination was simulated 30 times. Therefore, we have 30*4 = 120 instances run under each scenario. Specifically for FLAME, we categorized the data based on its median to adapt to this algorithm.

Figure \ref{fig:TPM_bias} illustrates the bias in ATT estimation across various scenarios. In all cases, incorporating our feature selection algorithm into these causal inference machine learning models significantly reduces both the bias and variance of ATT estimation. For instance, in Figure \ref{fig:TPM_bias}, the bias of BART without the proposed feature selection algorithm is higher compared to when our algorithm is applied. Similarly, for TMLE-SL, incorporating our feature selection algorithm results in TMLE-SL-FS exhibiting smaller bias and variance in ATT estimation.

%%%%%%%%%%%%Section 7%%%%%%%%%%%%
\section{Opioid Use Disorder Dataset}
%%%%%%%%%%%%Section 7%%%%%%%%%%%%

In this section, we validate our proposed framework using a large-scale US Healthcare dataset: the National Survey of Drug Use and Health (NSDUH) dataset. The same dataset has also been used in \cite{islam2021feature}. We leverage a similar experimental setup used in \cite{islam2021feature} with the same year from 2015 to 2019, which comprises 282,383 samples. It contains 54 covariates, covering education, income, and various health-related covariates. Here, we also investigate the effect of the binary indicator \texttt{udpypnr}, i.e., opioid use disorder in the past 12 months or more, on the binary outcome \texttt{suicide\_flag}, i.e. the participants ever thought about killing themselves in the past 12 months. 

By separating the individuals with recorded opioid misuse/dependency from those without, we identify 1,659 participants with documented opioid misuse or dependency (treatment group) and 208,819 participants with no history of opioid misuse or dependency (control group). Given the substantially larger size of the control group, we employ a bootstrap strategy 500 times, drawing 5,000 random samples from the control group in each iteration. All the tests are run on the 2.4 GHz Quad-Core Intel Core i5 CPU. 

To validate our proposed framework, we utilize the domain knowledge summarized in \cite{islam2021feature} by reviewing state-of-the-art literature. Specifically, we fit a regression model in the matched dataset using the features selected by the experts, and then we estimate the ATT by retrieving the coefficient of the treatment indicator. The precise meaning of each covariate abbreviation in the dataset and the expert choice of variables can be found by referring to \textbf{Online Supplement S1 Appendix I}.

During our evaluation, we utilize the ATT estimated from the (i) Sample ATT, when there is no feature selection (ii) Enh-ESVMS, which uses the SVM estimator and sigmoid function in our proposed three-stage framework (iii) Enh-ELRT, which uses a logistic regression estimator and tanh function in our proposed three-stage framework (iv) OAL (v) OAENet and (vi) BACR. We didn't include BCEE, Boruta\_T, Boruta\_Y and DWR since they are relatively time-consuming when processing this dataset compared to other algorithms.

\begin{figure}[htbp] 
    \centering
    \includegraphics[width=0.60\textwidth]{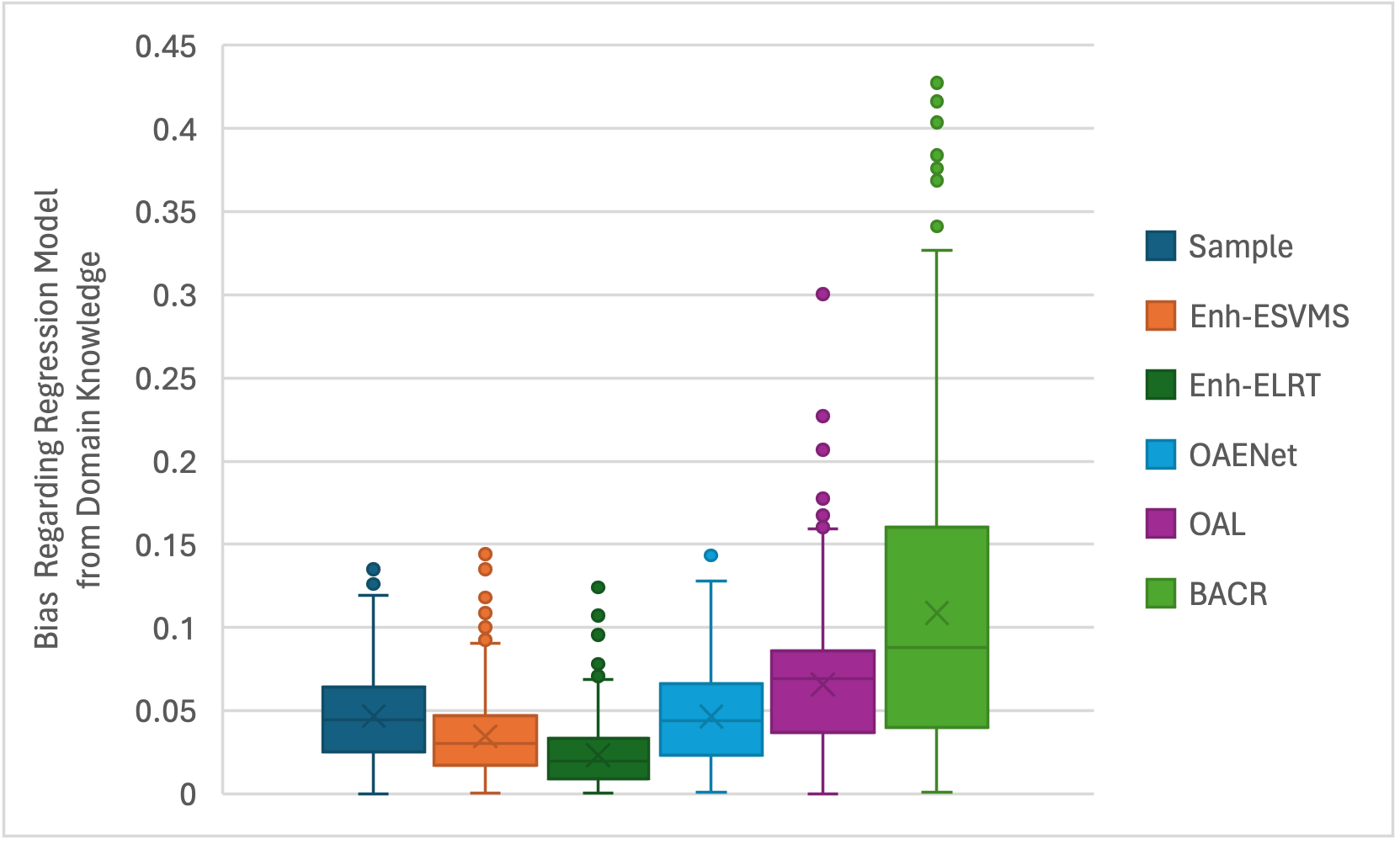}%
    \caption{Bias for each model regarding the estimated ATT from the domain knowledge. The ideal model/framework should have 0 bias.}
    \label{fig:oud_att}
\end{figure}

The average ATT estimated from the expert's domain knowledge is 0.0587, from (i) Sample ATT is 0.0144, from (ii) Enh-ESVMS is 0.0258, from (iii) Enh-ELRT is 0.0436, from (iv) OAL is 0.0812, from (v) OAENet is 0.0154, from (vi) BACR is 0.0179. Figure \ref{fig:oud_att} presents the bias of each model regarding ATT estimated from the expert's domain knowledge. Note that when estimating ATT, we only consider the features that are selected in at least 70\% of runs for each model. Since model (ii) and model (iii) provide lower bias and variance than the other benchmark models, we can claim that our proposed framework can efficiently reduce bias and variance of the estimation of the ATT in this dataset.

\begin{figure}[htbp] 
    \centering
    
    \resizebox{\textwidth}{2.2in}{
    \includegraphics[width=0.99\textwidth]{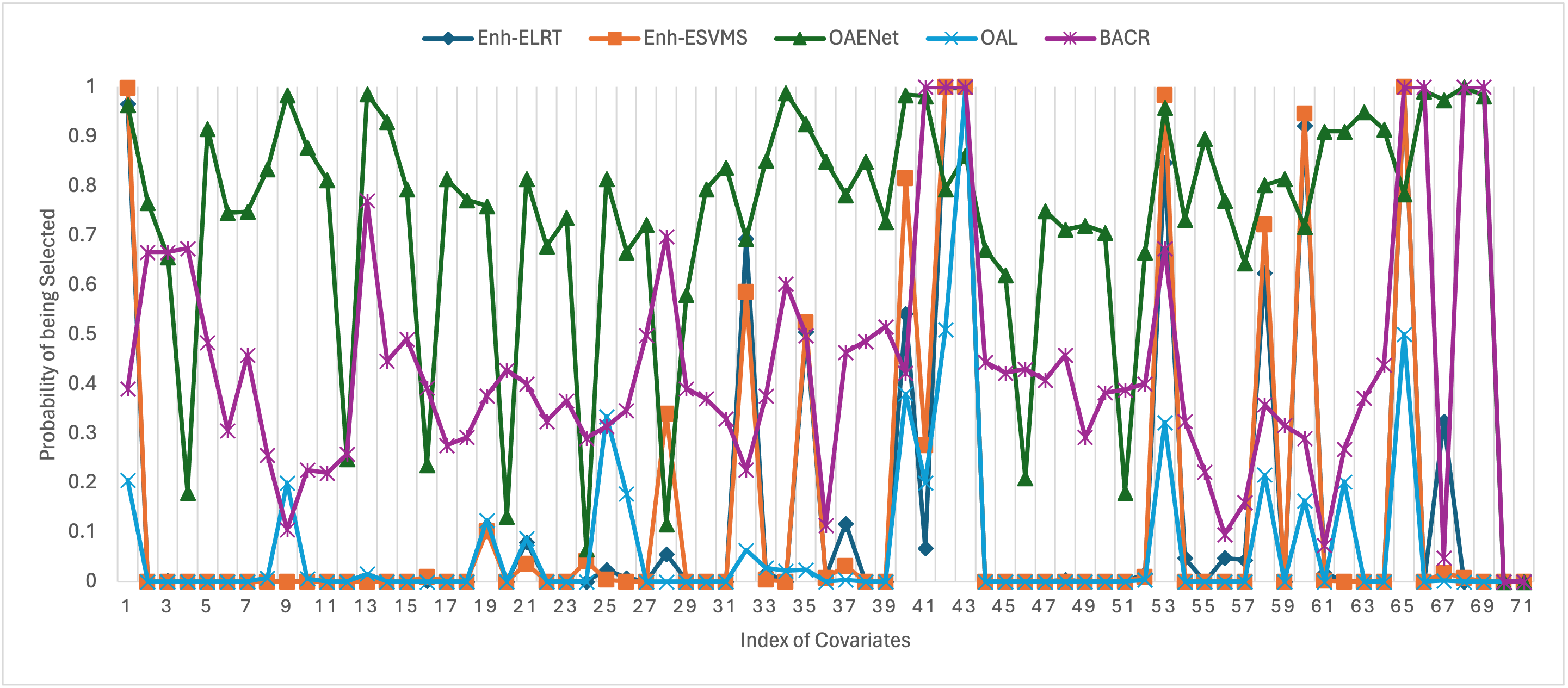}%
    }
    \caption{Probability of each variable selected from opioid use data out of 500 runs. The x-axis refers to the index of the covariates, while the y-axis indicates the probability of each variable selected.}
    \label{fig:oud_var_sel}
    
\end{figure}

We graphically present the probability of each variable selected by each model in Figure \ref{fig:oud_var_sel}. This figure illustrates the robustness of our proposed framework, as the selected set of variables almost remains consistent in different datasets drawn from the same data. This demonstrated the higher level of the oracle property achieved by our proposed framework in real-world datasets.

The average time spent for each iteration for (i) Enh-ESVMS is 30.701 seconds, for (ii) Enh-ELRT is 27.622 seconds, for (iii) OAL is 0.470 seconds, for (iv) OAENet is 28.434 seconds, for (v) BACR is 117.506 seconds. Our proposed models, however, exhibit superior performance in reducing bias in estimating ATT, while maintaining scalable computational efficiency to large-scale datasets.

The results of OUD dataset demonstrate the advantages of our proposed framework in feature selection tasks across the following aspects:
\\ \hspace*{1em} 1. \textbf{Reducing bias and variance}: As illustrated in Figure \ref{fig:oud_att}, our framework achieves lower bias and variance compared to other state-of-the-art feature selection methods in causal inference.
\\ \hspace*{1em} 2. \textbf{Consistency in selecting features}: Figure \ref{fig:oud_var_sel} highlights that our framework outperforms other algorithms in consistently selecting the same set of features across different draws from the original dataset, underscoring its stronger oracle property.

%%%%%%%%%%%%Section 8%%%%%%%%%%%%
\section{Conclusion}
%%%%%%%%%%%%Section 8%%%%%%%%%%%%

In this paper, we propose a novel feature selection framework for causal inference. The framework is designed to reduce selection bias and variance in causal analysis. To further improve model performance and achieve the oracle property, we develop an SVM estimator for the exposure model and design penalty smoothing functions, culminating in a robust three-stage feature selection framework. 

The experimental results from both synthetic and real-world data demonstrate the proposed framework's superiority over other state-of-the-art models in addressing selection bias, reducing variance, and achieving the oracle property. As a result, the feature selection models and algorithms introduced in this paper provide significant value for deriving robust causal insights from large-scale observational data across healthcare, social science, and engineering domains.

\if0\blind{
% \section*{Data Availability}
% The data and code can be retrieved from the GitHub link https://github.com/Tianyu-tj/Feature-Selection-Causal-Inference-Project/blob/main

\section*{Acknowledgements}
Financial support from the National Science Foundation (Award Number: 2047094) is greatly acknowledged.} \fi

\section*{Data availability statement}
This study uses two sets of experiments to support the findings. The first set consists of simulated data and section 6.1 provides a description to replicate the experiments. The second set of experiments uses the National Survey of Drug Use and Health (NSDUH) data from the Substance Abuse and Mental Health Services Administration (SAMHSA). This dataset is publicly available from SAMHSA website at “https://www.samhsa.gov/data/data-we-collect/nsduh-national-survey-drug-use-and-health”.

\bibliographystyle{chicago}
\spacingset{1}
\bibliography{IISE_Trans}

\begin{thebibliography}{}

\bibitem[\protect\citeauthoryear{Bald{\'e}, Yang, and Lefebvre}{Bald{\'e} et~al.}{2022}]{balde2022reader}
Bald{\'e}, I., Y.~A. Yang, and G.~Lefebvre (2022).
\newblock Reader reaction to “outcome-adaptive lasso: Variable selection for causal inference” by shortreed and ertefaie (2017).
\newblock {\em Biometrics\/}.

\bibitem[\protect\citeauthoryear{De~Luna, Waernbaum, and Richardson}{De~Luna et~al.}{2011}]{de2011covariate}
De~Luna, X., I.~Waernbaum, and T.~S. Richardson (2011).
\newblock Covariate selection for the nonparametric estimation of an average treatment effect.
\newblock {\em Biometrika\/}~{\em 98\/}(4), 861--875.

\bibitem[\protect\citeauthoryear{Ertefaie, Asgharian, and Stephens}{Ertefaie et~al.}{2018}]{ertefaie2018variable}
Ertefaie, A., M.~Asgharian, and D.~A. Stephens (2018).
\newblock Variable selection in causal inference using a simultaneous penalization method.
\newblock {\em Journal of Causal Inference\/}~{\em 6\/}(1).

\bibitem[\protect\citeauthoryear{Greenland}{Greenland}{2008}]{greenland2008invited}
Greenland, S. (2008).
\newblock Invited commentary: variable selection versus shrinkage in the control of multiple confounders.
\newblock {\em American journal of epidemiology\/}~{\em 167\/}(5), 523--529.

\bibitem[\protect\citeauthoryear{Gruber and van~der Laan}{Gruber and van~der Laan}{2010}]{gruber2010application}
Gruber, S. and M.~J. van~der Laan (2010).
\newblock An application of collaborative targeted maximum likelihood estimation in causal inference and genomics.
\newblock {\em The International Journal of Biostatistics\/}~{\em 6\/}(1).

\bibitem[\protect\citeauthoryear{Ho, Lim, Reza, and Xia}{Ho et~al.}{2017}]{ho2017om}
Ho, T.-H., N.~Lim, S.~Reza, and X.~Xia (2017).
\newblock Om forum—causal inference models in operations management.
\newblock {\em Manufacturing \& Service Operations Management\/}~{\em 19\/}(4), 509--525.

\bibitem[\protect\citeauthoryear{Iacus, King, and Porro}{Iacus et~al.}{2012}]{iacus2012causal}
Iacus, S.~M., G.~King, and G.~Porro (2012).
\newblock Causal inference without balance checking: Coarsened exact matching.
\newblock {\em Political analysis\/}~{\em 20\/}(1), 1--24.

\bibitem[\protect\citeauthoryear{Islam, Morshed, Young, and Noor-E-Alam}{Islam et~al.}{2019}]{islam2019robust}
Islam, M.~S., M.~S. Morshed, G.~J. Young, and M.~Noor-E-Alam (2019).
\newblock Robust policy evaluation from large-scale observational studies.
\newblock {\em PloS one\/}~{\em 14\/}(10), e0223360.

\bibitem[\protect\citeauthoryear{Islam, Shikalgar, and Noor-E-Alam}{Islam et~al.}{2024}]{islam2021feature}
Islam, M.~S., S.~Shikalgar, and M.~Noor-E-Alam (2024).
\newblock A two-stage feature selection approach for robust evaluation of treatment effects in high-dimensional observational data.
\newblock {\em IISE Transactions on Healthcare Systems Engineering, https://www.tandfonline.com/doi/full/10.1080/24725579.2024.2447715,\/}.

\bibitem[\protect\citeauthoryear{Ju, Gruber, Lendle, Chambaz, Franklin, Wyss, Schneeweiss, and van Der~Laan}{Ju et~al.}{2019}]{ju2019scalable}
Ju, C., S.~Gruber, S.~D. Lendle, A.~Chambaz, J.~M. Franklin, R.~Wyss, S.~Schneeweiss, and M.~J. van Der~Laan (2019).
\newblock Scalable collaborative targeted learning for high-dimensional data.
\newblock {\em Statistical methods in medical research\/}~{\em 28\/}(2), 532--554.

\bibitem[\protect\citeauthoryear{King and Nielsen}{King and Nielsen}{2019}]{king2019propensity}
King, G. and R.~Nielsen (2019).
\newblock Why propensity scores should not be used for matching.
\newblock {\em Political analysis\/}~{\em 27\/}(4), 435--454.

\bibitem[\protect\citeauthoryear{Kursa, Jankowski, and Rudnicki}{Kursa et~al.}{2010}]{kursa2010boruta}
Kursa, M.~B., A.~Jankowski, and W.~R. Rudnicki (2010).
\newblock Boruta--a system for feature selection.
\newblock {\em Fundamenta Informaticae\/}~{\em 101\/}(4), 271--285.

\bibitem[\protect\citeauthoryear{Lu}{Lu}{2020}]{lu2020feature}
Lu, R. (2020).
\newblock {\em Feature Selection for High Dimensional Causal Inference}.
\newblock Columbia University.

\bibitem[\protect\citeauthoryear{Pearl}{Pearl}{2012}]{pearl2012class}
Pearl, J. (2012).
\newblock On a class of bias-amplifying variables that endanger effect estimates.
\newblock {\em arXiv preprint arXiv:1203.3503\/}.

\bibitem[\protect\citeauthoryear{Roberts, Stewart, and Nielsen}{Roberts et~al.}{2015}]{roberts2015matching}
Roberts, M.~E., B.~M. Stewart, and R.~Nielsen (2015).
\newblock Matching methods for high-dimensional data with applications to text.
\newblock {\em Unpublished manuscript\/}.

\bibitem[\protect\citeauthoryear{Rosenbaum and Rubin}{Rosenbaum and Rubin}{1983}]{rosenbaum1983central}
Rosenbaum, P.~R. and D.~B. Rubin (1983).
\newblock The central role of the propensity score in observational studies for causal effects.
\newblock {\em Biometrika\/}~{\em 70\/}(1), 41--55.

\bibitem[\protect\citeauthoryear{Rotnitzky, Li, and Li}{Rotnitzky et~al.}{2010}]{rotnitzky2010note}
Rotnitzky, A., L.~Li, and X.~Li (2010).
\newblock A note on overadjustment in inverse probability weighted estimation.
\newblock {\em Biometrika\/}~{\em 97\/}(4), 997--1001.

\bibitem[\protect\citeauthoryear{Rubin}{Rubin}{1979}]{rubin1979using}
Rubin, D.~B. (1979).
\newblock Using multivariate matched sampling and regression adjustment to control bias in observational studies.
\newblock {\em Journal of the American Statistical Association\/}~{\em 74\/}(366a), 318--328.

\bibitem[\protect\citeauthoryear{Shortreed and Ertefaie}{Shortreed and Ertefaie}{2017}]{shortreed2017outcome}
Shortreed, S.~M. and A.~Ertefaie (2017).
\newblock Outcome-adaptive lasso: variable selection for causal inference.
\newblock {\em Biometrics\/}~{\em 73\/}(4), 1111--1122.

\bibitem[\protect\citeauthoryear{Stuart}{Stuart}{2010}]{stuart2010matching}
Stuart, E.~A. (2010).
\newblock Matching methods for causal inference: A review and a look forward.
\newblock {\em Statistical science: a review journal of the Institute of Mathematical Statistics\/}~{\em 25\/}(1), 1.

\bibitem[\protect\citeauthoryear{Talbot, Lefebvre, and Atherton}{Talbot et~al.}{2015}]{talbot2015bayesian}
Talbot, D., G.~Lefebvre, and J.~Atherton (2015).
\newblock The bayesian causal effect estimation algorithm.
\newblock {\em Journal of Causal Inference\/}~{\em 3\/}(2), 207--236.

\bibitem[\protect\citeauthoryear{VanderWeele}{VanderWeele}{2019}]{vanderweele2019principles}
VanderWeele, T.~J. (2019).
\newblock Principles of confounder selection.
\newblock {\em European journal of epidemiology\/}~{\em 34}, 211--219.

\bibitem[\protect\citeauthoryear{Wang, Parmigiani, and Dominici}{Wang et~al.}{2012}]{wang2012bayesian}
Wang, C., G.~Parmigiani, and F.~Dominici (2012).
\newblock Bayesian effect estimation accounting for adjustment uncertainty.
\newblock {\em Biometrics\/}~{\em 68\/}(3), 661--671.

\bibitem[\protect\citeauthoryear{Wang, Morucci, Awan, Liu, Roy, Rudin, and Volfovsky}{Wang et~al.}{2021}]{wang2021flame}
Wang, T., M.~Morucci, M.~U. Awan, Y.~Liu, S.~Roy, C.~Rudin, and A.~Volfovsky (2021).
\newblock Flame: A fast large-scale almost matching exactly approach to causal inference.
\newblock {\em The Journal of Machine Learning Research\/}~{\em 22\/}(1), 1477--1517.

\bibitem[\protect\citeauthoryear{Wooldridge}{Wooldridge}{2016}]{wooldridge2016should}
Wooldridge, J.~M. (2016).
\newblock Should instrumental variables be used as matching variables?
\newblock {\em Research in Economics\/}~{\em 70\/}(2), 232--237.

\bibitem[\protect\citeauthoryear{Yao, Chu, Li, Li, Gao, and Zhang}{Yao et~al.}{2021}]{yao2021survey}
Yao, L., Z.~Chu, S.~Li, Y.~Li, J.~Gao, and A.~Zhang (2021).
\newblock A survey on causal inference.
\newblock {\em ACM Transactions on Knowledge Discovery from Data (TKDD)\/}~{\em 15\/}(5), 1--46.

\bibitem[\protect\citeauthoryear{Zou}{Zou}{2006}]{zou2006adaptive}
Zou, H. (2006).
\newblock The adaptive lasso and its oracle properties.
\newblock {\em Journal of the American statistical association\/}~{\em 101\/}(476), 1418--1429.

\bibitem[\protect\citeauthoryear{Zou and Hastie}{Zou and Hastie}{2005}]{zou2005regularization}
Zou, H. and T.~Hastie (2005).
\newblock Regularization and variable selection via the elastic net.
\newblock {\em Journal of the royal statistical society: series B (statistical methodology)\/}~{\em 67\/}(2), 301--320.

\bibitem[\protect\citeauthoryear{Zou and Zhang}{Zou and Zhang}{2009}]{zou2009adaptive}
Zou, H. and H.~H. Zhang (2009).
\newblock On the adaptive elastic-net with a diverging number of parameters.
\newblock {\em Annals of statistics\/}~{\em 37\/}(4), 1733.

\end{thebibliography}
	
\end{document}